\documentclass[12pt,onecolumn,draftclsnofoot]{IEEEtran}

\usepackage[encapsulated]{CJK}
\usepackage{ucs}
\usepackage[utf8x]{inputenc}
\usepackage[cmex10]{amsmath}
\usepackage{amsmath,amssymb,amscd,bbm,amsthm,mathrsfs,dsfont}
\usepackage{algorithmic,algorithm}
\usepackage{mdwmath}
\usepackage{mdwtab}
\usepackage{bm,upgreek}
\usepackage{cite}
\usepackage{rotating,graphics,psfrag,epsfig}
\usepackage{array}
\usepackage{booktabs}
\usepackage{indentfirst}
\usepackage{subfigure}
\usepackage{lipsum,fancyhdr,lastpage,refcount}
\usepackage[hyphens]{url}
\usepackage{fixltx2e,dblfloatfix}
\usepackage{blindtext}
\usepackage[T1]{fontenc}
\usepackage{color}

\IEEEoverridecommandlockouts

\let\oldnl\nl
\newcommand{\nonl}{\renewcommand{\nl}{\let\nl\oldnl}}

\begin{document}

\title{\huge Information Freshness-Aware Task Offloading in Air-Ground Integrated Edge Computing Systems}

\author{\IEEEauthorblockN{Xianfu Chen, Celimuge Wu, Tao Chen, Zhi Liu, Honggang Zhang, Mehdi Bennis, Hang Liu, and Yusheng Ji}

\thanks{X. Chen and T. Chen are with the VTT Technical Research Centre of Finland, Finland (e-mail: \{xianfu.chen, tao.chen\}@vtt.fi). C. Wu is with the Graduate School of Informatics and Engineering, University of Electro-Communications, Tokyo, Japan (e-mail: celimuge@uec.ac.jp). Z. Liu is with the Department of Mathematical and Systems Engineering, Shizuoka University, Japan (e-mail: liu@ieee.org). H. Zhang is with the College of Information Science and Electronic Engineering (ISEE), Zhejiang University, Hangzhou, China (e-mail: honggangzhang@zju.edu.cn). M. Bennis is with the Centre for Wireless Communications, University of Oulu, Finland (e-mail: mehdi.bennis@oulu.fi). H. Liu is with the Department of Electrical Engineering and Computer Science, the Catholic University of America, USA (e-mail: liuh@cua.edu). Y. Ji is with the Information Systems Architecture Research Division, National Institute of Informatics, Tokyo, Japan (e-mail: kei@nii.ac.jp).}

\thanks{This work has been submitted to the IEEE for possible publication. Copyright may be transferred without notice, after which this version may no longer be accessible.}
}

\maketitle

\vspace{-1.6cm}
\begin{abstract}

This paper studies the problem of information freshness-aware task offloading in an air-ground integrated multi-access edge computing system, which is deployed by an infrastructure provider (InP).
A third-party real-time application service provider provides computing services to the subscribed mobile users (MUs) with the limited communication and computation resources from the InP based on a long-term business agreement.
Due to the dynamic characteristics, the interactions among the MUs are modelled by a non-cooperative stochastic game, in which the control policies are coupled and each MU aims to selfishly maximize its own expected long-term payoff.
To address the Nash equilibrium solutions, we propose that each MU behaves in accordance with the local system states and conjectures, based on which the stochastic game is transformed into a single-agent Markov decision process.
Moreover, we derive a novel online deep reinforcement learning (RL) scheme that adopts two separate double deep Q-networks for each MU to approximate the Q-factor and the post-decision Q-factor.
Using the proposed deep RL scheme, each MU in the system is able to make decisions without a priori statistical knowledge of dynamics.
Numerical experiments examine the potentials of the proposed scheme in balancing the age of information and the energy consumption.

\end{abstract}


\section{Introduction}
\label{intr}

By provisioning computation resources in close proximity to the mobile users (MUs), multi-access edge computing (MEC) is becoming one of the key technologies to mitigate the burden to resource-constrained mobile devices from the computation-intensive applications \cite{Mao17, Wang20}.
In an MEC system, the computation tasks of each MU can be processed locally at the mobile device or offloaded to a set of servers at the edge for remote execution.
Strategic computation offloading not only greatly improves the computation Quality-of-Experience (QoE) and Quality-of-Service (QoS), but also augments the capability of MUs for running a variety of emerging applications (e.g., virtual/augmented reality, mission-critical controls, etc.) \cite{Mao17}.
Recent years have witnessed a large body of research on designing computation offloading policies.
In \cite{Wang18}, Wang et al. proposed a Lagrangian duality method to minimize the total energy consumption in a computation latency constrained wireless powered multiuser MEC system.
In \cite{Liu17}, Liu et al. studied the power-delay tradeoff for an MEC system using the Lyapunov optimization technique.
In \cite{Apos20}, Apostolopoulos et al. analyzed the risk-seeking computation offloading behaviours of MUs in a multi-MEC server environment from a non-cooperative game-theoretic viewpoint.
In our priori work \cite{Chen19J}, the infinite time-horizon Markov decision process (MDP) framework was applied to formulate the problem of computation offloading for a representative MU in an ultra-dense radio access network (RAN) and to solve the optimal policy, we proposed the reinforcement learning (RL)-based schemes.
In \cite{He19}, He et al. identified the privacy vulnerability caused by the wireless communication feature of MEC-enabled Internet-of-Things (IoT), for which an effective computation offloading scheme based on the post-decision state learning algorithm was developed.

Offloading computation tasks from the mobile device of each MU to the edge servers relies on wireless data transmissions, which encounter high spatial-temporal communication uncertainties \cite{Sun19}.
In particular, the time-varying channel qualities due to the MU mobility in turn limit the overall computation performance \cite{Chen19, Abd19}.
Because of among others, the flexibility for convenient deployment and the desired line-of-sight (LOS) connections, unmanned aerial vehicles (UAVs) have been expected to play a significant role in advancing the future wireless networks \cite{Zeng16M, Moza19, Amor20}.
Integrating the UAV technology into a ground MEC system has been shown to be substantial.
In \cite{Hu19}, Hu et al. investigated an UAV-assisted MEC architecture, where an UAV acts as a computing server to help the MUs process the computation tasks or as a relay to offload the tasks to the access point for execution, and derived an alternating algorithm to optimize the weighted sum energy consumption.
In \cite{Shan20}, Shang and Liu implemented UAVs as aerial base stations (BSs) in an air-ground integrated MEC system and introduced a coordinate descent algorithm for the problem of total energy consumption minimization.
In \cite{Ashe19}, Asheralieva and Niyato presented a hierarchical game-theoretical and RL framework for computation offloading in an MEC network, where multiple service providers (SPs) install computing servers at both ground BSs and UAVs.

Despite the efforts focusing on technical implementation issues, the air-ground integrated MEC systems open up a sustainable business model in the mobile industry \cite{ETSI18}.
An infrastructure provider (InP) deploys the UAVs as the flying servers, complementary to the ground MEC system, which enables the third-party application SPs to provide the ubiquitous computing services to the subscribed MUs with computation requests.
At an UAV, the computation tasks of the MUs are executed in parallel by the created isolated virtual machines (VMs) \cite{Lian19}.
In this paper, we are primarily concerned with such a three-dimensional UAV-assisted MEC system operated by the InP in conjunction with a third-party application SP.
However, both technical and economic challenges arise.
On the one hand, most of the existing works (e.g., \cite{Hu19} and \cite{Shan20}) on computation offloading are based on a finite time-horizon.
It is expensive to repeatedly formulate the optimization problem in accordance with the dynamic characteristics of an air-ground integrated MEC system (i.e., the UAV and MU mobilities, the uncertain computation task arrivals, the unpredictable available communication and computation resources, etc.), which nevertheless fails to characterize the expected long-term computation offloading performance.
On the other hand, the economic issues of facilitating an air-ground integrated MEC system are overlooked (e.g., \cite{Ashe19}).
A long-term business agreement with the InP allows an SP to steer the computation requests to the edge computing facilities \cite{ETSI18}.
How to dynamically charge the computing services to the subscribed MUs for revenue maximization remains critical \cite{Wang20J}.

In contrast to the incurred delay, the QoE and QoS for many real-time applications are restricted by the information freshness of the computation outcomes \cite{Yate19, Yate20}, which adds another dimension of challenge to the computation offloading problem in an air-ground integrated MEC system.
In this paper, we employ the metric of age of information (AoI) to capture the information freshness \cite{Kaul11, Chen19A, Abde18}.
By definition, AoI is the amount of time elapsed since the outcome of the most recently scheduled computation was received \cite{Zhon19}.
It should be noted that there are a few related works studying the AoI under the context of edge computing.
In \cite{Zhon19}, Zhong et al. designed a greedy traffic scheduling policy to minimize the weighted sum of the average AoI over multiple MUs in edge applications.
In \cite{Xu20}, Xu et al. developed an analytical framework for an IoT system to investigate the effect of computing on the information freshness, which is in terms of peak AoI.
In \cite{Kuan20}, Kuang et al. studied the AoI for computation-intensive messages with MEC in status update scenarios.
The results of these works are limited to the ground MEC systems and hence are not widely applicable.

Different from the above literature, in this paper, we concentrate on the problem of information freshness-aware task offloading in an air-ground integrated MEC system.
More specifically, a third-party real-time application SP serves the subscribed MUs across the infinite time-horizon over a limited number of channels and computation resources from the InP.
Upon receiving the auction bids submitted by the non-cooperative MUs, the resource orchestrator (RO) of the SP manages the channel allocation through a Vickrey-Clarke-Groves (VCG) pricing mechanism \cite{Ji07}.
One major advantage of the VCG auction mechanism is that the dominant auction policy of an MU is to bid with the true valuation of the channels.
In addition, the VCG auction mechanism outperforms the generalized second-price auction for revenue produced to the SP \cite{Edel07}.
Consequently, each MU is able to not only process a computation task at the local mobile device, but also offload a computation task to the ground MEC server or to the UAV for remote execution via the channel won from the auction.
Sharing the same physical platform of an UAV for parallel execution among the MUs causes I/O interference, leading to computation rate reduction for each VM \cite{Lian19}.
%
%
%
In summary, the main contributions from this paper are threefold.
\begin{itemize}
  \item Taking into account the dynamics and the limited communication as well as computation resources in the air-ground integrated MEC system, we formulate the problem of information freshness-aware task offloading across the infinite time-horizon as a stochastic game under the framework of a multi-agent MDP, in which each MU aims to selfishly maximize its own expected long-term payoff from the interactions with other MUs.
      To the best of our knowledge, there does not exist a comprehensive study for the problem targeted in this paper.
  \item To avoid any private information exchange among the non-cooperative MUs, we propose that each MU behaves independently with the local conjectures, each of which preserves the payment to the SP from the channel auction and the experienced computation service rate at the UAV.
      The original stochastic game can hence be transformed into a single-agent MDP.
  \item Without a priori statistical knowledge of dynamics and to deal with the huge local state space faced by each MU, we put forward a novel online deep RL scheme leveraging the double deep Q-network (DQN) \cite{Hass16}.
      The proposed deep RL scheme maintains for each MU two separate DQNs to approximate, respectively, the Q-factor and the post-decision Q-factor, similar to a deep advantage actor-critic (A2C) architecture \cite{Mnih16}.
\end{itemize}

The remainder of this paper is organized as follows.
In the next section, we describe the air-ground integrated MEC system and the assumptions used throughout this paper.
In Section \ref{prob}, we formulate the information freshness-aware task offloading as a stochastic game among the non-cooperative MUs and discuss the general best-response solution.
In Section \ref{probSolv}, we elaborate how each MU plays the stochastic game with the local conjectures and propose an online deep RL scheme to address the optimal control policy.
In Section \ref{simu}, we provide numerical experiments under various settings to compare the performance from our scheme with other baselines.
Finally, we draw the conclusions in Section \ref{conc}.
For convenience, Table \ref{tabl0} summarizes the major notations of this paper.
\begin{table*}
  \caption{Major notations used in the paper.}\label{tabl0}
        \begin{center}
        \begin{tabular}{|c|l||c|l|}
              \hline
  \tiny Notation                                              & \tiny Description
 &\tiny Notation                                              & \tiny Description                                                           \\\hline
  \tiny $B$/$\mathcal{B}$                                     & \tiny number/set of BSs
 &\tiny $\mathcal{L}_b$                                       & \tiny set of locations covered by BS $b$                                    \\\hline
  \tiny $\mathcal{K}$                                         & \tiny set of MUs
 &\tiny $\delta$                                              & \tiny time duration of one decision epoch                                   \\\hline
  \tiny $\mathcal{C}$                                         & \tiny set of channels
 &\tiny $H$                                                   & \tiny flying altitude of UAV                                                \\\hline
  \tiny $\eta$                                                & \tiny bandwidth of a channel
 &\tiny $\bm\beta_k$, $\bm\beta_k^j$                          & \tiny auction bid of MU $k$                                                 \\\hline
  \tiny $\nu_k$, $\nu_k^j$                                    & \tiny true valuation of MU $k$
 &\tiny $\mathbf{N}_k$, $\mathbf{N}_k^j$                      & \tiny channel demand profile of MU $k$                                      \\\hline
  \tiny $\bm\rho_k^j$                                         & \tiny channel allocation vector of MU $k$
 &\tiny $\varphi_k$, $\varphi_k^j$                            & \tiny channel allocation variable of MU $k$                                 \\\hline
  \tiny $\bm\phi$, $\bm\phi^k$                                & \tiny auction winner determination vector
 &\tiny $\tau_k$, $\tau_k^j$                                  & \tiny payment of MU $k$                                                     \\\hline
  \tiny $L_{(\mathrm{v}), k}$, $L_{(\mathrm{v}), k}^j$        & \tiny location of UAV
 &\tiny $L_{(\mathrm{m}), k}$, $L_{(\mathrm{m}), k}^j$        & \tiny location of MU $k$                                                    \\\hline
  \tiny $\lambda$                                             & \tiny task generation probability
 &\tiny $\zeta_k^j$                                           & \tiny task arrival indicator of MU $k$                                      \\\hline
  \tiny $D_{(\max)}$                                          & \tiny number of input data packets of a task
 &\tiny $\mu$                                                 & \tiny number of bits of an input data packet                                \\\hline
  \tiny $\vartheta$                                           & \tiny required CPU cycles per bit
 &\tiny $\varrho$                                             & \tiny CPU-cycle frequency of an MU                                          \\\hline
  \tiny $\Delta$                                              & \tiny number of epochs to locally finish a task
 &\tiny $\varsigma$                                           & \tiny effective switched capacitance                                        \\\hline
  \tiny $X_k$, $X_k^j$                                        & \tiny task offloading decision of MU $k$
 &\tiny $R_k$, $R_k^j$                                        & \tiny packet scheduling decision of MU $k$                                  \\\hline
  \tiny $I_k$, $I_k^j$                                        & \tiny association state of MU $k$
 &\tiny $T_k$, $T_k^j$                                        & \tiny arrival epoch index of buffered task of MU $k$                        \\\hline
  \tiny $\tilde{\delta}_k^j$                                  & \tiny exact transmission time of MU $k$ in epoch $j$
 &\tiny $\xi$                                                 & \tiny handover delay                                                        \\\hline
  \tiny $A_{(\max)}$                                          & \tiny upper limit of AoI
 &\tiny $\gamma$                                              & \tiny discounted factor                                                     \\\hline
  \tiny $G_{b, k}^j$/$G_{(\mathrm{v}), k}^j$                  & \tiny channel power gain between MU $k$ and BS $b$/UAV
 &\tiny $P_{(\max)}$                                          & \tiny maximum transmit power                                                \\\hline
  \tiny $W_{(\mathrm{m}), k}$, $W_{(\mathrm{m}), k}^j$        & \tiny local CPU state of MU $k$
 &\tiny $W_{(\mathrm{v}), k}$, $W_{(\mathrm{v}), k}^j$        & \tiny remote processing state of MU $k$                                     \\\hline
  \tiny $D_k$, $D_k^j$                                        & \tiny local transmitter state of MU $k$
 &\tiny $\chi^j$                                              & \tiny computation service rate                                              \\\hline
  \tiny $F_k$, $F_k^j$                                        & \tiny total local energy consumption of MU $k$
 &\tiny $F_{(\mathrm{m}), k}^j$                               & \tiny CPU energy consumption of MU $k$                                      \\\hline
  \tiny $F_{(\mathrm{s}), k}^j$, $F_{(\mathrm{v}), k}^j$      & \tiny transmit energy consumption of MU $k$
 &\tiny $A_k$, $A_k^j$                                        & \tiny AoI of MU $k$                                                         \\\hline
  \tiny $\ell_k$                                              & \tiny payoff function of MU $k$
 &\tiny $u_k$                                                 & \tiny utility function of MU $k$                                            \\\hline
  \tiny $\varpi_k$                                            & \tiny AoI weight for MU $k$
 &\tiny $\omega_k$                                            & \tiny total energy consumption weight of MU $k$                             \\\hline
  \tiny $\mathbf{S}$, $\mathbf{S}^j$                          & \tiny global system state
 &\tiny $\mathbf{S}_k$, $\mathbf{S}_k^j$                      & \tiny local system state of MU $k$                                          \\\hline
  \tiny $\widehat{\mathbf{S}}_k$, $\widehat{\mathbf{S}}_k^j$  & \tiny local state of MU $k$
 &\tiny $\mathbf{O}_k$, $\mathbf{O}_k^j$                      & \tiny local conjecture of MU $k$                                            \\\hline
  \tiny $\widetilde{\mathbf{S}}_k$                            & \tiny local post-decision state of MU $k$
 &\tiny $\bm\pi$, $\bm\pi^*$                                  & \tiny joint control policy                                                  \\\hline
  \tiny $\bm\pi_k$, $\bm\pi_k^*$                              & \tiny control policy of MU $k$
 &\tiny $\pi_{(\mathrm{c}), k}$, $\pi_{(\mathrm{c}), k}^*$    & \tiny channel auction policy of MU $k$                                      \\\hline
  \tiny $\pi_{(\mathrm{t}), k}$, $\pi_{(\mathrm{t}), k}^*$    & \tiny task offloading policy of MU $k$
 &\tiny $\pi_{(\mathrm{p}), k}$, $\pi_{(\mathrm{p}), k}^*$    & \tiny packet scheduling policy of MU $k$                                    \\\hline
  \tiny $V_k$                                                 & \tiny expected long-term payoff of MU $k$
 &\tiny $Q_k$                                                 & \tiny Q-factor of MU $k$                                                    \\\hline
  \tiny $\widetilde{Q}_k$                                     & \tiny post-decision Q-factor of MU $k$
 &\tiny $\bm\theta_k$, $\bm\theta_k^j$, $\bm\theta_k^{j, -}$  & \tiny parameters associated with the DQN-I of MU $k$                        \\\hline
  \tiny $\tilde{\bm\theta}_k$, $\tilde{\bm\theta}_k^j$        & \tiny parameters associated with the DQN-II of MU $k$
 &\tiny $\mathcal{M}_k^j$                                     & \tiny replay memory of MU $k$                                               \\\hline
  \tiny $\mathcal{Y}_k^j$                                     & \tiny mini-batch of MU $k$
 &\tiny $\epsilon$                                            & \tiny exploration probability
              \\\hline
        \end{tabular}
    \end{center}
\end{table*}

\section{System Descriptions and Assumptions}
\label{sysm}

\begin{figure}[t]
  \centering
  \includegraphics[width=20pc]{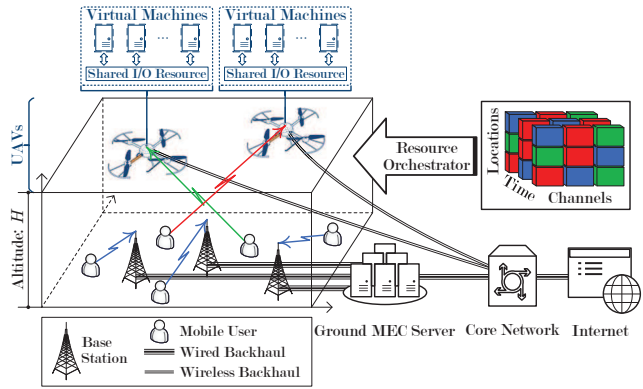}
  \caption{Illustration of an air-ground integrated multi-access edge computing (MEC) system, where the unmanned aerial vehicles (UAVs) are deployed as the flying servers.
  A third-party real-time application service provider (SP) serves the subscribed mobile users (MUs) with sporadic computation requests.
  The resource orchestrator of the SP is responsible for allocating a limited number of channels to the MUs across the decision epochs based on the submitted auction bids.}
  \label{systMode}
\end{figure}

In this paper, we assume an InP deploys a three-dimensional UAV-assisted MEC system as shown in Fig. \ref{systMode}, where the ground MEC server and the UAVs jointly provide computing capability at the edge.
A set $\mathcal{B} = \{1, 2, \cdots, B\}$ of BSs in the RAN are connected via the wired backhaul to the resource-rich ground MEC server, while each UAV works as a parallel computing server.
Based on a long-term business agreement with the InP, a third-party real-time application SP serves over the system a set $\mathcal{K}$ of subscribed MUs with sporadic computation requests.
The UAVs fly in the air at a fixed altitude of $H$ (in meters) \footnote{This work assumes that the power of the UAVs is supplied by laser charging \cite{Liu19}. Hence the UAVs are able to operate for the long run.
Under the RL framework \cite{Rich98}, the proposed study in this paper can be straightforwardly applied to the episodic case in which an episode is defined as the maximum UAV operation time, if an UAV needs to land on the ground for battery recharging \cite{Abd19}.
}.
We choose a finite set $\mathcal{L}$ of locations (i.e., small two-dimensional non-overlapping areas) to denote both the service region covered by the RAN and the region of the UAVs mapped vertically from the air to the ground.
A location or small area can be characterized by uniform wireless communication conditions \cite{Abd19, Chen19}.
Let $\mathcal{L}_b$ denote the locations covered by an BS $b \in \mathcal{B}$.
For any two BSs $b$ and $b' \in \mathcal{B} \setminus \{b\}$, we assume that $\mathcal{L}_b \cap \mathcal{L}_{b'} = \emptyset$.
Thus, $\mathcal{L} = \cup_{b \in \mathcal{B}} \mathcal{L}_b$.
The geographical topology of the BSs is represented by a two-tuple graph $\langle\mathcal{B}, \mathcal{E}\rangle$, where $\mathcal{E} = \{e_{b, b'}: b, b' \in \mathcal{B}, b \neq b'\}$ with each $e_{b, b'}$ being equal to $1$ if BSs $b$ and $b'$ are neighbours, and $0$, otherwise.
The infinite time-horizon is divided into discrete decision epochs, each of which is with equal duration $\delta$ (in seconds) and indexed by an integer $j \in \mathds{N}_+$.
To ease the following analysis, we concentrate on the air-ground integrated MEC system with a single UAV without loss of generality.
The results in this paper can be easily extended to the multi-UAV scenario by expanding the dimension of the task offloading decision-makings.

\subsection{VCG-based Channel Auction}

In the service region, we assume that the UAV and the MUs move at the same speed following a Markov mobility model\footnote{Other mobility models \cite{Wan13, Xi19}, including changing the flying altitude within the operating region \cite{Amer20}, can also be applied but do not affect the proposed scheme in this paper.
We leave the UAV trajectory optimization for part of our future investigation.}.
Let $L_{(\mathrm{v})}^j \in \mathcal{L}$ and $L_{(\mathrm{m}), k}^j \in \mathcal{L}$ denote, respectively, the mapped ground location of the UAV and the location of each MU $k \in \mathcal{K}$ during a decision epoch $j$.
The computation task arrivals at the MUs are assumed to be independent and identically distributed sequences of Bernoulli random variables with a common parameter $\lambda \in [0, 1]$.
More specifically, we denote by $\zeta_k^j \in \{0, 1\}$ the task arrival indicator for an MU $k$, that is, $\zeta_k^j = 1$ if a computation task is generated at MU $k$ at the beginning of a decision epoch $j$ and otherwise, $\zeta_k^j = 0$.
Then, $\mathbb{P}(\zeta_k^j = 1) = 1 - \mathbb{P}(\zeta_k^j = 0) = \lambda$, $\forall k \in \mathcal{K}$, where $\mathbb{P}(\cdot)$ means the probability of the occurrence of an event.
Each MU $k$ employs a pre-processing buffer to temporarily store a computation task.
It is reasonable for an incoming task with newer arrival time to replace an old task in the buffer since a newer computation task is always with fresher information.
We assume that a computation task is composed of $D_{(\max)}$ input data packets and each data packet contains $\mu$ bits.
We let $\vartheta$ represent the number of CPU cycles required to accomplish one bit of a computation task.
A computation task can be either computed locally at the mobile device of the MU or executed remotely (at the ground MEC server or the UAV).
We let $X_k^j \in \mathcal{X} = \{0, 1, 2, 3\}$ denote the computation offloading decision of MU $k$ at each decision epoch $j$, where $X_k^j = 1$, $X_k^j = 2$ and $X_k^j = 3$ indicate that the task in the pre-processing buffer is scheduled to be processed by the local CPU, executed by the ground MEC server and offloaded to the UAV for execution, respectively, while $X_k^j = 0$ means that the task is not scheduled for computation.
The RO of the SP manages a finite set $\mathcal{C}$ of non-overlapping orthogonal channels, each of which is with the same bandwidth $\eta$ (in Hz).
In order to upload the input data packets of a scheduled computation task for remote execution, an MU competes with other non-cooperative MUs in the system for the limited channel access opportunities using an VCG auction mechanism.

Specifically, at the beginning of each decision epoch $j$, each MU $k \in \mathcal{K}$ submits to the RO an auction bid given by a vector $\bm\beta_k^j = (\nu_k^j, \mathbf{N}_k^j)$, where $\nu_k^j$ is the true valuation over $\mathbf{N}_k^j = (N_{(\mathrm{s}), k}^j, N_{(\mathrm{v}), k}^j)$ with $N_{(\mathrm{s}), k}^j$ and $N_{(\mathrm{v}), k}^j$ being the numbers of demanded channels for transmitting the input data packets to the ground MEC server and the UAV.
Let $\bm\rho_k^j = (\rho_{k, c}^j: c \in \mathcal{C})$ be the channel allocation vector for MU $k$ during epoch $j$, where $\rho_{k, c}^j$ equals $1$ if a channel $c \in \mathcal{C}$ is allocated to MU $k$ during epoch $j$ and $0$, otherwise.
We consider
\begin{align}
        \left(\sum_{k \in \mathcal{K}_{(\mathrm{s}), b}^j} \rho_{k, c}^j\right) \cdot
        \left(\sum_{k \in \mathcal{K}_{(\mathrm{s}), b'}^j} \rho_{k, c}^j\right)
 & =    0, \mbox{if } e_{b, b'} = 1, \forall e_{b, b'} \in  \mathcal{E}, \forall c \in \mathcal{C};     \label{c1}\\
        \left(\sum_{k \in \underset{b \in \mathcal{B}}{\cup} \mathcal{K}_{(\mathrm{s}), b}^j}
        \rho_{k, c}^j\right) \cdot
        \left(\sum_{k \in \mathcal{K}_{(\mathrm{v})}^j}
        \rho_{k, c}^j\right)
 & =    0, \forall c \in \mathcal{C};                                                                   \label{c2}\\
        \sum_{k \in \mathcal{K}_{(\mathrm{s}), b}^j} \rho_{k, c}^j \leq 1, \forall b
 & \in  \mathcal{B}, \forall c \in \mathcal{C};                                                         \label{c3}\\
        \sum_{k \in \mathcal{K}_{(\mathrm{v})}^j} \rho_{k, c}^j
 & \leq 1, \forall c \in \mathcal{C};                                                                   \label{c4}\\
        \sum_{c \in \mathcal{C}} \rho_{k, c}^j
 & \leq 1, \forall k \in \mathcal{K},                                                                   \label{c5}
\end{align}
for the centralized channel allocation at the RO during each decision epoch $j$ to ensure that
\begin{enumerate}
  \item[\emph{1)}] a channel cannot be allocated simultaneously to the MUs covered by two adjacent BSs if the MUs transmit the input data packets to the ground MEC server;
  \item[\emph{2)}] a channel cannot be shared between the data transmissions to the ground MEC server and the UAV; and
  \item[\emph{3)}] an MU can be assigned at most one channel, and in the coverage of an BS, a channel can be assigned to at most one MU.
\end{enumerate}
In above, $\mathcal{K}_{(\mathrm{s}), b}^j = \{k: k \in \mathcal{K}, L_{(\mathrm{m}), k}^j \in \mathcal{L}_b, N_{(\mathrm{s}), k}^j > 0\}$, $\forall b \in \mathcal{B}$, while $\mathcal{K}_{(\mathrm{v})}^j = \{k: k \in \mathcal{K}, N_{(\mathrm{v}), k}^j > 0\}$.
The independent data transmissions can be hence guaranteed among the MUs.
Obviously, we have the following
\begin{align}\label{c6}
  N_{(\mathrm{s}), k}^j + N_{(\mathrm{v}), k}^j \leq 1, \forall k \in \mathcal{K}, \forall j,
\end{align}
that constrains the design of an auction bid.

We denote $\bm\phi^j = (\phi_k^j: k \in \mathcal{K})$ as the winner determination in the channel auction at a decision epoch $j$, where $\phi_k^j = 1$ if an MU $k \in \mathcal{K}$ wins the channel auction while $\phi_k^j = 0$ indicates that no channel is allocated to MU $k$ during the epoch.
The RO calculates $\bm\phi^j$ according to
\begin{equation}\label{chanSche}
  \begin{array}{cl}
                  & \bm\phi^j = \underset{\bm\phi}{\arg\max} \displaystyle\sum\limits_{k \in \mathcal{K}} \phi_k \cdot \nu_k^j    \\
    \mathrm{s.t.} & \mbox{constraints (\ref{c1}), (\ref{c2}), (\ref{c3}), (\ref{c4}) and (\ref{c5})};                                   \\
                  &\!\!
                  \begin{array}{r@{~}l}
                    \displaystyle\sum_{k \in \mathcal{K}_{(\mathrm{s}), b}^j} \varphi_k^j &= \phi_k \cdot N_{(\mathrm{s}), k}^j,
                        \forall b \in \mathcal{B}, \forall k \in \mathcal{K}; \\
                    \displaystyle\sum_{k \in \mathcal{K}_{(\mathrm{v})}^j} \varphi_k^j    &= \phi_k \cdot N_{(\mathrm{v}), k}^j, \forall k \in \mathcal{K},
                  \end{array}
  \end{array}
\end{equation}
where $\bm\phi = (\phi_k \in \{0, 1\}: k \in \mathcal{K})$ and $\varphi_k^j = \sum_{c \in \mathcal{C}} \rho_{k, c}^j$ is a channel allocation variable that equals $1$ if MU $k$ is assigned a channel during the decision epoch and $0$, otherwise.
For consistency, we also rewrite $\varphi_k^j$ as $\varphi_k(\bm\beta^j)$, where $\bm\beta^j = (\bm\beta_k^j, \bm\beta_{- k}^j)$ with $- k$ denoting all the other MUs in $\mathcal{K}$ without the presence of MU $k$.
Moreover, the payment for MU $k$ to the SP, which is incurred from accessing the allocated channel, is calculated to be
\begin{align}\label{paymCalc}
   \tau_k^j =
   \max\limits_{\bm\phi_{-k}} \displaystyle\sum\limits_{\kappa \in \mathcal{K} \setminus \{k\}} \phi_{\kappa} \cdot \nu_{\kappa}^j
            - \displaystyle\sum\limits_{\kappa \in \mathcal{K} \setminus \{k\}} \phi_{\kappa}^j \cdot \nu_{\kappa}^j.
\end{align}
It has been known that the VCG-based channel auction satisfies the economic properties: \emph{1)} computational efficiency; \emph{2)} individual rationality; and \emph{3)} truthfulness \cite{Chen19}.

\subsection{Computation and Communication Models}

The UAV complements the ground MEC system with the computation resource from the air.
By strategically offloading the computation tasks to the ground MEC server or the UAV for remote execution, the MUs can expect a significantly optimized computation experience.
Let $T_k^j \in \mathds{N}$ be the arrival epoch index of the computation task waiting in the pre-processing buffer of an MU $k \in \mathcal{K}$ at the beginning of a decision epoch $j$.
By default, we set $T_k^j = 0$ if the pre-processing buffer is empty.

\subsubsection{Local Computation}

When a computation task is scheduled for processing locally at the mobile device of an MU $k \in \mathcal{K}$ during a decision epoch $j$, i.e., $X_k^j = 1$, the number of required epochs can be calculated as $\Delta = \lceil (D_{(\max)} \cdot \mu \cdot \vartheta) / (\delta \cdot \varrho) \rceil$, where $\lceil \cdot \rceil$ means the ceiling function and we assume that the local CPU of an MU operates at frequency $\varrho$ (in Hz).

We describe by $W_{(\mathrm{m}), k}^j \in \{0, 1, \cdots, \Delta\}$ the local CPU state of each MU $k \in \mathcal{K}$ at the beginning of each decision epoch $j$, which is the number of remaining epochs to accomplish the scheduled computation task.
In particular, $W_{(\mathrm{m}), k}^j = 0$ indicates that the local CPU is idle and is available for a new task from epoch $j$.
The energy (in Joules) consumed by local CPU during epoch $j$ is then given by
\begin{align}\label{locaEnerCons}
   F_{(\mathrm{m}), k}^j =
   \left\{\!\!
    \begin{array}{l@{}l}
     0,                                         & \mbox{ for } W_{(\mathrm{m}), k}^j = 0;\\
     \varsigma \cdot \left(D_{(\max)} \cdot \mu \cdot \vartheta -
        \left(\Delta - 1\right) \cdot \delta \cdot \varrho\right) \cdot
        (\varrho)^2,                            & \mbox{ for } W_{(\mathrm{m}), k}^j = 1;\\
     \varsigma \cdot \delta \cdot (\varrho)^3,  & \mbox{ for } W_{(\mathrm{m}), k}^j > 1,
    \end{array}
   \right.
\end{align}
where $\varsigma$ is the effective switched capacitance that depends on the chip architecture of the mobile device of an MU \cite{Burd96}.

\subsubsection{Remote Execution}

To upload the input data packets under remote execution, an MU has to be first associated to the RAN (via one of the BSs depending on the geographical locations of the MU) or with the UAV until the task is finished.
Let $I_k^j \in \mathcal{B} \cup \{B + 1\}$ be the association state of each MU $k \in \mathcal{K}$ at the beginning of a decision epoch $j$, namely, $I_k^j = b \in \mathcal{B}$ if MU $k$ is associated with an BS $b$ and if MU $k$ is associated with the UAV, $I_k^j = B + 1$.
If no computation task is being scheduled during epoch $j$, the association state of MU $k$ is set according to
\begin{align}
  I_k^j =
  \left\{\!\!
  \begin{array}{l@{~}l}
    I_k^{j - 1}, & \mbox{for } I_k^{j - 1} = B + 1;                                                                 \\
    b,           & \mbox{for } I_k^{j - 1} \in \mathcal{B} \mbox{ and } L_{(\mathrm{m}), k}^j \in \mathcal{L}_b.
  \end{array}
  \right.
\end{align}
When $I_k^{j + 1} \neq I_k^j$, $\forall j$, a handover is triggered \cite{Chen19J}.
We assume that the energy consumption during the occurrence of one handover is negligible for MU $k$ but the handover delay is $\xi$ (in seconds).
The exact transmission time of MU $k$ during an epoch $j$ can be written as
\begin{align}
    \tilde{\delta}_k^j = \delta - \xi \cdot \mathbf{1}_{\left\{I_k^{j + 1} \neq I_k^j\right\}},
\end{align}
where the indicator function $\mathbf{1}_{\{i\}}$ equals $1$ if the condition $i$ is met and $0$ otherwise.
Let $D_k^j \in \mathcal{D} = \{0, 1, \cdots, D_{(\max)}\}$ denote the local transmitter state of MU $k$ at the beginning of each decision epoch $j$, which is defined as the number of input data packets left at the transmitter for uploading.
Let $R_k^j$ be the number of input data packets that are scheduled for transmissions during epoch $j$, the transmitter state of MU $k$ then evolves to
\begin{align}
    D_k^{j + 1} = D_k^j - \varphi_k^j \cdot R_k^j.
\end{align}
During a decision epoch $j$, each MU $k$ experiences the average channel power gains $G_{b, k}^j = g_{(\mathrm{s})}(L_{(\mathrm{m}), k}^j)$ for the link to each BS $b$ and $G_{(\mathrm{v}), k}^j = g_{(\mathrm{v})}(L_{(\mathrm{m}), k}^j, L_{(\mathrm{v})}^j)$ for the link to the UAV. 
Notice that $0 \leq R_k^j \leq \min\{D_k^j, R_{(\max), k}^j\}$, where $R_{(\max), k}^j$ is jointly determined by the channel gain during a decision epoch $j$, the transmission time and the maximum transmit power $P_{(\max)}$ at the MUs.

At the beginning of a decision epoch $j$, if an MU $k \in \mathcal{K}$ schedules the computation task in the pre-processing buffer for execution at the ground MEC server, namely, $X_k^j = 2$.
During the subsequent decision epochs, all the input data packets need to be uploaded via the allocated channels from the VCG auctions over the RAN.
When $L_{(\mathrm{m}), k}^j \in \mathcal{L}_b$, $b \in \mathcal{B}$, the energy consumed for reliably transmitting $\varphi_k^j \cdot R_k^j$ input data packets of the computation task to the ground MEC server is calculated as
\begin{align}\label{tranEner}
   F_{(\mathrm{s}), k}^j =
   \frac{\tilde{\delta}_k^j \cdot \eta \cdot \sigma^2} {G_{b, k}^j} \cdot \left(2^{\frac{\varphi_k^j \cdot \left(\mu \cdot R_k^j\right)}
   {\eta \cdot \tilde{\delta}_k^j}} - 1\right),
\end{align}
where $\sigma^2$ is the noise power spectral density.
In this paper, we assume that the ground MEC server is of rich computation resource and accordingly, the task execution delay is ignored.
Further, the time consumption (by an BS or the UAV) for sending the computation outcome back to the MU is negligible, due to the fact that the computation outcome is in general much smaller than the input data packets \cite{Chen16}.

In this paper, we assume that once all the input data packets of a computation task are received up to a current decision epoch, the UAV starts to execute from the beginning of next epoch, when the VMs are created for the MUs \cite{Lian19}.
If an MU $k \in \mathcal{K}$ decides to upload the computation task to the UAV for execution (i.e., $X_k^j = 3$),
the energy consumption of transmitting $\varphi_k^j \cdot R_k^j$ input data packets to the UAV during an epoch $j$ turns to be
\begin{align}\label{tranEnerUAV}
   F_{(\mathrm{v}), k}^j =
   \frac{\tilde{\delta}_k^j \cdot \eta \cdot \sigma^2} {G_{(\mathrm{v}), k}^j} \cdot \left(2^{\frac{\varphi_k^j \cdot \left(\mu \cdot R_k^j\right)}
   {\eta \cdot \tilde{\delta}_k^j}} - 1\right).
\end{align}
Let $\breve{\mathcal{K}}_{(\mathrm{v})}^j$ represent the set of MUs, whose computation tasks are being simultaneously executed at the UAV during a decision epoch $j$.
Denote by $\chi_0$ the computation service rate (in bits per second) of an VM created by the UAV given that the task is executed in isolation, the degraded computation service rate of an MU $k \in \breve{\mathcal{K}}_{(\mathrm{v})}^j$ is modeled as $\chi^j = \chi_0 \cdot (1 + \varepsilon)^{1 - |\breve{\mathcal{K}}_{(\mathrm{v})}^j|}$, where $|\cdot|$ denotes the cardinality of a set and $\varepsilon \in \mathds{R}_+$ is a factor specifying the percentage of reduction in the computation service rate of an VM when multiplexed with another VM at the UAV.
We then update the remote processing state of MU $k$ by $W_{(\mathrm{v}), k}^{j + 1} = \max\{W_{(\mathrm{v}), k}^j - \chi^j \cdot \delta, 0\}$, where $W_{(\mathrm{v}), k}^j$ quantifies the amount of input data bits remaining at the UAV at the beginning of an epoch $j$.

\subsection{AoI Evolution}

For each MU $k \in \mathcal{K}$ in the air-ground integrated MEC system, we define the AoI as the difference between the current time of receiving the outcome of the latest scheduled computation task and the corresponding task arrival time.
The AoI metric depicts the information freshness for MU $k$ from the task computing process.
Let $A_k^j$ denote the AoI of MU $k$ at each decision epoch $j$.
In line with the discussions, an arriving computation task can be either computed at the local CPU of MU $k$, or executed remotely at the ground MEC server or the UAV.
Depending on whether or not the computation outcomes are received during an epoch $j$, the AoI evolution of each MU $k$ can be analysed in three cases.
\begin{enumerate}
\item[\emph{1)}] When there is no computation outcome received at MU $k$ during decision epoch $j$, the AoI increases linearly according to $A_k^{j + 1} = A_k^j + \delta$.
\item[\emph{2)}] If MU $k$ receives only one computation outcome during decision epoch $j$, the AoI is then updated to be
    \begin{align}\label{AoI1}
       &  A_k^{j + 1} =                                                                                             \\
       &  \left\{\!\!
            \begin{array}{l@{}l}
              \left(j - T_{(\mathrm{m}), k}^j - \Delta + 1\right) \cdot \delta +
                      \dfrac{D_{(\max)} \cdot \mu \cdot \vartheta}{\varrho},
                    & \mbox{ for } W_{(\mathrm{m}), k}^j = 1, D_k^j = 0 \mbox{ and } W_{(\mathrm{v}), k}^j = 0;     \\
              \left(j - T_{(\mathrm{s}), k}^j + 1\right) \cdot \delta,
                    & \mbox{ for } W_{(\mathrm{m}), k}^j = 0, D_k^j > 0 \mbox{ and } W_{(\mathrm{v}), k}^j = 0;     \\
              \left(j - T_{(\mathrm{v}), k}^j\right) \cdot \delta + \dfrac{W_{(\mathrm{v}), k}^j}{\chi^j},
                    & \mbox{ for } W_{(\mathrm{m}), k}^j = 0, D_k^j = 0 \mbox{ and } W_{(\mathrm{v}), k}^j > 0,
            \end{array}
          \right.                                                                                                   \nonumber
    \end{align}
    where $T_{(\mathrm{m}), k}^j$, $T_{(\mathrm{s}), k}^j$ and $T_{(\mathrm{v}), k}^j$ are, respectively, the arrival epoch indices of the tasks computed at the local CPU, the ground MEC server and the UAV.
\item[\emph{3)}] The AoI evolution of MU $k$ can be expressed as
    \begin{align}\label{AoI2}
        & A_k^{j + 1} =                                                                                     \\
        & \left\{\!\!
            \begin{array}{l@{}l}
              \left(j - T_{(\mathrm{s}), k}^j + 1\right) \cdot \delta,
                  & \mbox{ for } D_k^j > 0, W_{(\mathrm{v}), k}^j = 0
                    \mbox{ and } T_{(\mathrm{s}), k}^j > T_{(\mathrm{m}), k}^j;                             \\
              \left(j - T_{(\mathrm{v}), k}^j\right) \cdot \delta + \dfrac{W_{(\mathrm{v}), k}^j}{\chi^j},
                  & \mbox{ for } D_k^j = 0, W_{(\mathrm{v}), k}^j > 0
                    \mbox{ and } T_{(\mathrm{v}), k}^j > T_{(\mathrm{m}), k}^j;                             \\
              \left(j - T_{(\mathrm{m}), k}^j - \Delta + 1\right) \cdot \delta +
                    \dfrac{D_{(\max)} \cdot \mu \cdot \vartheta}{\varrho},
                  & \mbox{ otherwise},
            \end{array}
          \right.                                                                                           \nonumber
    \end{align}
    when two computation outcomes arrive during decision epoch $j$.
\end{enumerate}
In this paper, the value $A_k^j$ of AoI is initialized to be $A_k^1 = 0$ and up-limited by $A_{(\max)}$ for each MU $k$.
When $A_k^j = A_{(\max)}$, it means that the information from the computation outcomes is too stale for MU $k$.

\section{Game-Theoretic Problem Statement}
\label{prob}

In this section, we first formulate the problem of information freshness-aware task offloading across the infinite time-horizon from a game-theoretic perspective and then discuss the best-response solution. 

\subsection{Stochastic Game Formulation}

During each decision epoch $j$, the local system state of an MU $k \in \mathcal{K}$ can be described by $\mathbf{S}_k^j = (L_{(\mathrm{v})}^j, L_{(\mathrm{m}), k}^j, \mathbf{1}_{\{T_k^j > 0\}},$ $I_k^j, W_{(\mathrm{m}), k}^j, W_{(\mathrm{v}), k}^j, D_k^j, A_k^j) \in \mathcal{S}$, where $\mathcal{S}$ denotes a common local state space for all MUs in the considered air-ground integrated MEC system.
Then $\mathbf{S}^j = (\mathbf{S}_k^j, \mathbf{S}_{- k}^j) \in \mathcal{S}^{|\mathcal{K}|}$ characterizes the global system state during decision epoch $j$.
Let $\bm\pi_k = (\pi_{(\mathrm{c}), k}, \pi_{(\mathrm{t}), k}, \pi_{(\mathrm{p}), k})$ denote the stationary control policy of MU $k$, where $\pi_{(\mathrm{c}), k}$, $\pi_{(\mathrm{t}), k}$ and $\pi_{(\mathrm{p}), k}$ are the channel auction, the task offloading and the packet scheduling policies, respectively.
It is worth noting that $\pi_{(\mathrm{p}), k}$ is MU-specified and dependent on $\mathbf{S}_k^j$ only.
The joint control policy of all MUs can be given by $\bm\pi = (\bm\pi_k, \bm\pi_{- k})$.
When deploying $\bm\pi_k$, MU $k$ observes $\mathbf{S}^j$ at the beginning of each decision epoch $j$ and accordingly, submits the channel auction bid as well as makes the decisions of computation task offloading and input data packet scheduling, that is, $\bm\pi_k(\mathbf{S}^j) = (\pi_{(\mathrm{c}), k}(\mathbf{S}^j), \pi_{(\mathrm{t}), k}(\mathbf{S}_k^j),$ $\pi_{(\mathrm{p}), k}(\mathbf{S}_k^j)) = (\bm\beta_k^j, X_k^j, R_k^j)$.
We define an immediate payoff function\footnote{To stabilize the training of the proposed scheme in this paper, we choose an exponential function for the definition of a payoff utility, whose value does not dramatically diverge.
Moreover, the exponential function has been well fitted to the generic quantitative relationship between the QoE and the QoS \cite{Fied10}.}
for MU $k$ by
\begin{align}\label{utilFunc}
     \ell_k\!\left(\mathbf{S}^j, \left(\varphi_k^j, X_k^j, R_k^j\right)\right)
   = u_k\!\left(\mathbf{S}^j, \left(\varphi_k^j, X_k^j, R_k^j\right)\right)
   - \tau_k^j,
\end{align}
in which the utility function $u_k(\mathbf{S}^j, (\varphi_k^j, X_k^j, R_k^j)) = \varpi_k \cdot \exp(- A_k^j) + \omega_k \cdot \exp(- F_k^j)$ measures the satisfaction of information freshness and total local energy consumption $F_k^j =$ $F_{(\mathrm{m}), k}^j + F_{(\mathrm{s}), k}^j + F_{(\mathrm{v}), k}^j$ during each decision epoch $j$, $\varphi_k^j =$ $\varphi_k(\bm\pi_{(\mathrm{c})}(\mathbf{S}^j))$ with $\bm\pi_{(\mathrm{c})} = (\pi_{(\mathrm{c}), k}, \bm\pi_{(\mathrm{c}), - k})$ representing the joint channel auction policy, while $\varpi_k \in \mathds{R}_+$ and $\omega_k \in \mathds{R}_+$ are the weighting constants.

It is easy to verify that the randomness hidden in a sequence of the global system state realizations over the infinite time-horizon $\{\mathbf{S}^j: j \in \mathds{N}_+\}$ is Markovian with the controlled state transition probability given by
\begin{align}\label{statTranProb}
 & \mathbb{P}\!\left(\mathbf{S}^{j + 1} | \mathbf{S}^j,
        \left(\bm\varphi\!\left(\bm\pi_{(\mathrm{c})}\!\left(\mathbf{S}^j\right)\right),
        \bm\pi_{(\mathrm{t})}\!\left(\mathbf{S}^j\right),
        \bm\pi_{(\mathrm{p})}\!\left(\mathbf{S}^j\right)\right)\right)
 = \mathbb{P}\!\left(L_{(\mathrm{v})}^{j + 1} | L_{(\mathrm{v})}^j\right) \cdot
        \prod_{k \in \mathcal{K}} \mathbb{P}\!\left(L_{(\mathrm{m}), k}^{j + 1} | L_{(\mathrm{m}), k}^j\right) \cdot    \\
 & \mathbb{P}\!\left(\left(\mathbf{1}_{\{T_k^{j + 1} > 0\}}, I_k^{j + 1},
        W_{(\mathrm{m}), k}^{j + 1}, W_{(\mathrm{v}), k}^{j + 1}, D_k^{j + 1}, A_k^{j + 1}\right) |
        \left(\mathbf{1}_{\{T_k^j > 0\}}, I_k^j, W_{(\mathrm{m}), k}^j,
        W_{(\mathrm{v}), k}^j, D_k^j, A_k^j\right), \bm\pi_k\!\left(\mathbf{S}^j\right)\right),                         \nonumber
\end{align}
where $\bm\varphi(\bm\pi_{(\mathrm{c})}(\mathbf{S}^j)) = (\varphi_k(\bm\pi_{(\mathrm{c})}(\mathbf{S}^j)),$ $\bm\varphi_{- k}(\bm\pi_{(\mathrm{c})}(\mathbf{S}^j)))$ is the global channel allocation by the RO, while $\bm\pi_{(\mathrm{t})} = (\pi_{(\mathrm{t}), k}, \bm\pi_{(\mathrm{t}), - k})$ and $\bm\pi_{(\mathrm{p})} = (\pi_{(\mathrm{p}), k}, \bm\pi_{(\mathrm{p}), - k})$ are the joint task offloading and the joint packet scheduling policies, respectively.
Given the control policy $\bm\pi_k$ by each MU $k \in \mathcal{K}$ and an initial global system state $\mathbf{S} = (\mathbf{S}_k = (L_{(\mathrm{v})}, L_{(\mathrm{m}), k},$ $\mathbf{1}_{\{T_k > 0\}}, I_k, W_{(\mathrm{m}), k}, W_{(\mathrm{v}), k}, D_k, A_k): k \in \mathcal{K}) \in \mathcal{S}^{|\mathcal{K}|}$, we express the expected long-term discounted payoff function of MU $k$ as below
\begin{align}\label{statValu}
    V_k(\mathbf{S}, \bm\pi) =
    (1 - \gamma) \cdot \textsf{E}_{\bm\pi}\!\!\left[\sum_{j = 1}^\infty (\gamma)^{j  - 1} \cdot
    \ell_k\!\left(\mathbf{S}^j, \left(\varphi_k^j, X_k^j, R_k^j\right)\right) | \mathbf{S}^1 = \mathbf{S}\right],
\end{align}
where $\gamma \in [0, 1)$ is the discount factor and the expectation $\textsf{E}_{\bm\pi}[\cdot]$ is taken over different decision-makings under different global system states following the joint control policy $\bm\pi$ across the discrete decision epochs.
When $\gamma$ approaches $1$, (\ref{statValu}) well approximates the expected long-term un-discounted payoff\footnote{The non-cooperative interactions among MUs in the system result in that the control policies, $\bm\pi_k$, $\forall k \in \mathcal{K}$, are not unichain.
Therefore, the Markovian system is non-ergodic, due to which we continue using (\ref{statValu}) as the optimization goal for each MU.} \cite{Adel08}.
$V_k(\mathbf{S}, \bm\pi)$ in (\ref{statValu}) is also termed as the state-value function of the global system state $\mathbf{S}$ under the joint control policy $\bm\pi$ \cite{Rich98}.

Due to the limited number of channels managed by the RO, the shared I/O resource at the physical platform of the UAV and the dynamic characteristics of the air-ground integrated MEC system, we formulate the problem of information freshness-aware task offloading among the competing MUs over the infinite time-horizon as a non-cooperative stochastic game, in which $|\mathcal{K}|$ MUs are the players and there are a set $\mathcal{S}^{|\mathcal{K}|}$ of global system states and a collection of control policies $\{\bm\pi_k: \forall k \in \mathcal{K}\}$.
The objective of each MU $k$ in the stochastic game is to device a best-response control policy $\bm\pi_k^* = (\pi_{(\mathrm{c}), k}^*, \pi_{(\mathrm{t}), k}^*, \pi_{(\mathrm{p}), k}^*)$ that maximizes its own $V_k(\mathbf{S}, \bm\pi)$ for an any given global system state $\mathbf{S} \in \mathcal{S}^{|\mathcal{K}|}$, which can be formulated as
\begin{equation}\label{bestResp}
  \bm\pi_k^* = \underset{\bm\pi_k}{\arg\max}~ V_k(\mathbf{S}, \bm\pi), \forall \mathbf{S} \in \mathcal{S}^{|\mathcal{K}|}.
\end{equation}
A Nash equilibrium (NE) describes the rational behaviours of the MUs in a stochastic game.
Specifically, an NE is a tuple of control policies $\langle \bm\pi_k^*: k \in \mathcal{K}\rangle$, where each $\bm\pi_k^*$ of an MU $k$ is the best response to $\bm\pi_{-k}^*$.
Theorem 1 ensures the existence of an NE in our formulated game.

\emph{Theorem 1.}
For the $|\mathcal{K}|$-player stochastic game with expected long-term discounted payoffs, there always exists an NE in stationary control policies \cite{Fink64}.

For brevity, define $V_k(\mathbf{S}) = V_k(\mathbf{S}, \bm\pi_k^*, \bm\pi_{-k}^*)$ as the optimal state-value function, $\forall k \in \mathcal{K}$, $\forall \mathbf{S} \in \mathcal{S}^{|\mathcal{K}|}$.
From (\ref{statValu}), we can easily observe that the expected long-term payoff of an MU $k \in \mathcal{K}$ depends on information of not only the global system states across the time-horizon but also the joint control policy $\bm\pi$.
In other words, the decision-makings from all MUs are coupled in the stochastic game.

\subsection{Best-Response Approach}

Suppose that in the formulated stochastic game, the global system state information over the infinite time-horizon is perfectly known to all MUs and all MUs behave following the NE control policy profile $\bm\pi^* = (\bm\pi_k^*, \bm\pi_{- k}^*)$, the best-response of each MU $k \in \mathcal{K}$ under a global system state $\mathbf{S} \in \mathcal{S}^{|\mathcal{K}|}$ can then be given in the form of
\begin{align}\label{VFunc}
 &   V_k(\mathbf{S})
   = \max\limits_{\bm\pi_k(\mathbf{S})}\! \bigg\{ (1 - \gamma) \cdot
        \ell_k\!\left(\mathbf{S}, \varphi_k\!\left(\pi_{(\mathrm{c}), k}(\mathbf{S}),
        \bm\pi_{(\mathrm{c}), -k}^*(\mathbf{S})\right),
        \pi_{(\mathrm{t}), k}(\mathbf{S}_k), \pi_{(\mathrm{p}), k}(\mathbf{S}_k)\right) + \gamma \cdot                          \\
 &   \sum_{\mathbf{S}' \in \mathcal{S}^{|\mathcal{K}|}}
        \mathbb{P}\!\left(\mathbf{S}' | \mathbf{S}, \left(\bm\varphi\!\left(\pi_{(\mathrm{c}), k}(\mathbf{S}),
        \bm\pi_{(\mathrm{c}), -k}^*(\mathbf{S})\right), \left(\pi_{(\mathrm{t}), k}(\mathbf{S}),
        \bm\pi_{(\mathrm{t}), -k}^*(\mathbf{S})\right), \left(\pi_{(\mathrm{p}), k}(\mathbf{S}_k),
        \bm\pi_{(\mathrm{p}), -k}^*(\mathbf{S}_{-k})\right)\right)\right) \cdot V_k(\mathbf{S}') \bigg\},            \nonumber
\end{align}
where $\mathbf{S}' = (\mathbf{S}_k' = (L_{(\mathrm{v})}', L_{(\mathrm{m}), k}', \mathbf{1}_{\{T_k' > 0\}}, I_k', W_{(\mathrm{m}), k}', W_{(\mathrm{v}), k}', D_k',$ $A_k'): k \in \mathcal{K})$ is the consequent global system state.
We note that in order to operate in the NE, all MUs have to have a priori the statistical knowledge of global dynamics (i.e., (\ref{statTranProb})), which is prohibited for a non-cooperative system.

\section{Deep RL with Local Conjectures}
\label{probSolv}

In this section, we shall elaborate on how the MUs play the non-cooperative stochastic game only with limited local information.
Our aim is to develop an online deep RL scheme to approach the NE control policy with the local conjectures from the interactions among the competing MUs.

\subsection{Local Conjectures}
\label{optiSolu}

During the competitive interactions in the stochastic game, it is challenging for each MU $k \in \mathcal{K}$ to obtain the private system state information at other MUs.
On the other hand, the coupling of the decision-makings by the non-cooperative MUs exists in the channel auction and the remote task execution at the UAV.
From the viewpoint of an MU $k$, the payment $\tau_k^j$ to the SP in the channel auction and the computation service rate\footnote{It is straightforward that during each epoch $j$, the computation service rate $\chi^j$ of an MU $k \in \breve{\mathcal{K}}_{(\mathrm{v})}^j$ can be estimated locally with $W_{(\mathrm{v}), k}^j$, $W_{(\mathrm{v}), k}^{j + 1}$ and the time consumption by the respective VM at the UAV.} $\chi^j$ at each decision epoch $j$ are realized under $\mathbf{S}_{- k}^j$.
In our previous works \cite{Chen19, Chen1801}, an abstract game was constructed to approximate the stochastic game with a bounded performance regret.
However, the approximation bound highly depends on the abstraction mechanisms \cite{Kroe16}.
Instead, in this paper, we allow each MU $k$ to conjecture $\mathbf{S}^{j + 1}$ during the next decision epoch $j + 1$ as $\widehat{\mathbf{S}}_k^{j + 1} = (\mathbf{S}_k^{j + 1}, \mathbf{O}_k^{j + 1})$, where $\mathbf{O}_k^{j + 1} = (\tau_k^j, \chi^j) \in \mathcal{O}_k$ with $\mathcal{O}_k$ being the finite space\footnote{From the assumptions made throughout the paper, the payments and the computation service rates take discrete values. Therefore, the finite space $\mathcal{O}_k$ is sufficiently large.} of all possible local conjectures.
Now we are able to transform (\ref{statValu}) into
\begin{align}\label{apprValu}
   V_k\!\left(\widehat{\mathbf{S}}_k, \bm\pi\right) =
   (1 - \gamma) \cdot \textsf{E}_{\bm\pi}\!\!\left[\sum_{j = 1}^\infty (\gamma)^{j  - 1} \cdot
   \ell_k\!\left(\mathbf{S}^j, \left(\varphi_k^j, X_k^j, R_k^j\right)\right) |
   \widehat{\mathbf{S}}_k^1 = \widehat{\mathbf{S}}_k\right],
\end{align}
where $\widehat{\mathbf{S}}_k = (\mathbf{S}_k, \mathbf{O}_k) \in \widehat{\mathcal{S}}_k = \mathcal{S} \times \mathcal{O}_k$ with $\mathbf{O}_k$ being the initial local conjecture of $\mathbf{S}_{- k}$\footnote{The conjecture $\mathbf{O}_k$ of each MU $k \in \mathcal{K}$ at decision epoch $j = 1$ can be initialized to be, for example, $(0, 0)$ as in numerical simulations.}, while $\bm\pi$ hereinafter refers to the conjecture based joint control policy.
Each MU $k$ then switches to maximize $V_k(\widehat{\mathbf{S}}_k, \bm\pi)$, $\forall \widehat{\mathbf{S}}_k \in \widehat{\mathcal{S}}_k$, which is basically a single-agent MDP.
With a slight abuse of notation, we let $V_k(\widehat{\mathbf{S}}_k) = V_k(\widehat{\mathbf{S}}_k, \bm\pi^*)$, $\forall k \in \mathcal{K}$, where $\bm\pi^*$ is the best-response control policy profile of all MUs with local conjectures and the Bellman's optimality equation is given by
\begin{align}\label{VFuncn}
 &   V_k\!\left(\widehat{\mathbf{S}}_k\right)
   = \max\limits_{\bm\pi_k\!\left(\widehat{\mathbf{S}}_k\right)}\! \left\{ (1 - \gamma) \cdot
        \ell_k\!\left(\mathbf{S}, \varphi_k\!\left(\pi_{(\mathrm{c}), k}\!\left(\widehat{\mathbf{S}}_k\right),
        \bm\pi_{(\mathrm{c}), -k}^*\!\left(\widehat{\mathbf{S}}_{- k}\right)\right),
        \pi_{(\mathrm{t}), k}\!\left(\widehat{\mathbf{S}}_k\right), \pi_{(\mathrm{p}), k}\!\left(\mathbf{S}_k\right)\right)
        \vphantom{\sum_{\widehat{\mathbf{S}}_k' \in \widehat{\mathcal{S}}_k}}\right. +                                    \nonumber\\
 & \left.\gamma \cdot \sum_{\widehat{\mathbf{S}}_k' \in \widehat{\mathcal{S}}_k}
        \mathbb{P}\!\left(\widehat{\mathbf{S}}_k' | \widehat{\mathbf{S}}_k, \left(\varphi_k\!\left(\pi_{(\mathrm{c}), k}\!\left(\widehat{\mathbf{S}}_k\right),
        \bm\pi_{(\mathrm{c}), -k}^*\!\left(\widehat{\mathbf{S}}_{- k}\right)\right),
        \pi_{(\mathrm{t}), k}\!\left(\widehat{\mathbf{S}}_k\right),
        \pi_{(\mathrm{p}), k}\!\left(\mathbf{S}_k\right)\right)\right) \cdot
        V_k\!\left(\widehat{\mathbf{S}}_k'\right) \right\}.
\end{align}

With the observation of local state $\widehat{\mathbf{S}}_k \in \widehat{\mathcal{S}}_k$ at the beginning of a current decision epoch, each MU $k \in \mathcal{K}$ in the system submits an optimal auction bid $\pi_{(\mathrm{c}), k}^*(\widehat{\mathbf{S}}_k) = (\nu_k, \mathbf{N}_k)$ to the RO, which includes a true valuation $\nu_k$ of occupying $\mathbf{N}_k =$ $(N_{(\mathrm{s}), k}, N_{(\mathrm{v}), k})$ channels.
We have Theorem 2 that provides the optimal configuration of $(\nu_k, \mathbf{N}_k)$.

\emph{Theorem 2:}
When all MUs in the system follow the best-response control policy profile $\bm\pi^*$ based on the local conjectures, each MU $k \in \mathcal{K}$ announces at the beginning of a current decision epoch to the RO the channel demands
\begin{align}
    N_{(\mathrm{s}), k} & = z_k \cdot \mathbf{1}_{\left\{\pi_{(\mathrm{t}), k}^*\!\left(\widehat{\mathbf{S}}_k\right) = 2\right\}},  \label{auctBid1_0}\\
    N_{(\mathrm{v}), k} & = z_k \cdot \mathbf{1}_{\left\{\pi_{(\mathrm{t}), k}^*\!\left(\widehat{\mathbf{S}}_k\right) = 3\right\}},  \label{auctBid1_1}
\end{align}
together with the true valuation being specified as
\begin{align}\label{auctBid0}
     \nu_k
 & = u_k\!\left(\mathbf{S}, \left(z_k, \pi_{(\mathrm{t}), k}^*\!\left(\widehat{\mathbf{S}}_k\right),
     \pi_{(\mathrm{p}), k}^*(\mathbf{S}_k)\right)\right)                                               \nonumber\\
 & + \frac{\gamma}{1 - \gamma} \cdot \sum_{\widehat{\mathbf{S}}_k'\in \widehat{\mathcal{S}}_k}
     \mathbb{P}\!\left(\widehat{\mathbf{S}}_k' | \widehat{\mathbf{S}}_k,
     \left(z_k, \pi_{(\mathrm{t}), k}^*\!\left(\widehat{\mathbf{S}}_k\right), \pi_{(\mathrm{p}), k}^*(\mathbf{S}_k)\right)\right)
     \cdot V_k\!\left(\widehat{\mathbf{S}}_k'\right),
\end{align}
where $z_k \in \{0, 1\}$ is the preference of winning one channel from the VCG auction centralized at the RO and satisfies
\begin{align}\label{auctBid2}
     z_k
 & = \underset{z \in \{0, 1\}}{\arg\max} \bigg\{ (1 - \gamma) \cdot \ell_k\!\left(\mathbf{S}, \left(z,
     \pi_{(\mathrm{t}), k}^*\!\left(\widehat{\mathbf{S}}_k\right), \pi_{(\mathrm{p}), k}^*\!\left(\mathbf{S}_k\right)\right)\right) \nonumber\\
 & + \gamma \cdot \sum_{\widehat{\mathbf{S}}_k' \in \widehat{\mathcal{S}}_k}
   \mathbb{P}\!\left(\widehat{\mathbf{S}}_k' | \widehat{\mathbf{S}}_k,
   \left(z, \pi_{(\mathrm{t}), k}^*\!\left(\widehat{\mathbf{S}}_k\right),
   \pi_{(\mathrm{p}), k}^*\!\left(\mathbf{S}_k\right)\right)\right) \cdot
   V_k\!\left(\widehat{\mathbf{S}}_k'\right)\bigg\}.
\end{align}

\emph{Proof:}
The conjecture based best-response control policy $\bm\pi_k^*$ of each MU $k \in \mathcal{K}$ in the air-ground integrated MEC system consists of the channel auction policy $\pi_{(\mathrm{c}), k}^*$, the task offloading policy $\pi_{(\mathrm{t}), k}^*$ and the packet scheduling policy $\pi_{(\mathrm{p}), k}^*$.
We hence restructure (\ref{VFuncn}) as
\begin{align}\label{BellEqui}
 &   \pi_{(\mathrm{c}), k}^*\!\left(\widehat{\mathbf{S}}_k\right)
   = \underset{\bm\beta_k}{\arg\max}\! \left\{\ell_k\!\left(\mathbf{S},
     \left(\varphi_k\!\left(\bm\beta_k, \bm\pi_{(\mathrm{c}), -k}^*\!\left(\widehat{\mathbf{S}}_{- k}\right)\right),
     \pi_{(\mathrm{t}), k}^*\!\left(\widehat{\mathbf{S}}_k\right), \pi_{(\mathrm{p}), k}^*\!\left(\mathbf{S}_k\right)\right)\right)
     \vphantom{\sum_{\widehat{\mathbf{S}}_k' \in \widehat{\mathcal{S}}_k}}\right. +                                           \nonumber\\
 &   \left.\frac{\gamma}{1 - \gamma} \cdot \sum_{\widehat{\mathbf{S}}_k' \in \widehat{\mathcal{S}}_k}
     \mathbb{P}\!\left(\widehat{\mathbf{S}}_k' | \widehat{\mathbf{S}}_k,
     \left(\varphi_k\!\left(\bm\beta_k, \bm\pi_{(\mathrm{c}), -k}^*\!\left(\widehat{\mathbf{S}}_{- k}\right)\right),
     \pi_{(\mathrm{t}), k}^*\!\left(\widehat{\mathbf{S}}_k\right),
     \pi_{(\mathrm{p}), k}^*\!\left(\mathbf{S}_k\right)\right)\right) \cdot
     V_k\!\left(\widehat{\mathbf{S}}_k'\right)\right\},
\end{align}
$\forall \widehat{\mathbf{S}}_k \in \widehat{\mathcal{S}}_k$, where $\bm\beta_k = \pi_{(\mathrm{c}), k}(\widehat{\mathbf{S}}_k)$.
From the rules of winner determination in (\ref{chanSche}) as well as payment calculation in (\ref{paymCalc}), the optimal channel auction policy for MU $k$ is to bid truthfully across the decision epochs according to (\ref{auctBid1_0}), (\ref{auctBid1_1}) and (\ref{auctBid0}).
\hfill$\Box$

Without knowing the statistical dynamic characteristics of the local states and the structure of the payment function in VCG auction, it yet remains technically challenging for an MU in the air-ground integrated MEC system to come up with an optimal bid configured by (\ref{auctBid1_0}), (\ref{auctBid1_1}) and (\ref{auctBid0}) at the beginning of each decision epoch.

\subsection{Post-Decision Q-Factor}

In order to remove the obstacle for the calculations of an optimal auction bid at the beginning of each decision epoch, we introduce a local post-decision state (as in \cite{Salo08, Chen1801, Chen18S}) for the MUs in the considered air-ground integrated MEC system.
At each current decision epoch in the infinite time-horizon, the local post-decision state of an MU $k \in \mathcal{K}$ is defined as $\widetilde{\mathbf{S}}_k = (L_{(\mathrm{v})}, L_{(\mathrm{m}), k}, \mathbf{1}_{\{T_k > 0\}}, I_k, W_{(\mathrm{m}), k}, W_{(\mathrm{v}), k}, \widetilde{D}_k,$ $A_k, \mathbf{O}_k) \in \widehat{\mathcal{S}}_k$ by intentionally letting $\widetilde{D}_k = D_k - \varphi_k(\bm\beta) \cdot R_k$, where $\bm\beta =$ $(\bm\beta_k,$ $\bm\beta_{- k})$.
The local post-decision state in this paper can be interpreted as a local intermediate state right after the input data packet transmissions but before the transition into the next local state.
Accordingly, the probability of the transition from $\widehat{\mathbf{S}}_k$ to $\widehat{\mathbf{S}}_k'$ under a conjecture based joint control policy $\bm\pi$ can be expressed as
\begin{align}\label{locaPartStatTran}
     \mathbb{P}\!\left(\widehat{\mathbf{S}}_k' | \widehat{\mathbf{S}}_k, (\varphi_k(\bm\beta), X_k, R_k)\right) =
     \mathbb{P}\!\left(\widetilde{\mathbf{S}}_k | \widehat{\mathbf{S}}_k, (\varphi_k(\bm\beta), X_k, R_k)\right) \cdot
     \mathbb{P}\!\left(\widehat{\mathbf{S}}_k' | \widetilde{\mathbf{S}}_k, (\varphi_k(\bm\beta), X_k, R_k)\right),
\end{align}
where it admits $\mathbb{P}(\widetilde{\mathbf{S}}_k | \widehat{\mathbf{S}}_k, (\varphi_k(\bm\beta), X_k, R_k)) = 1$.

For each MU $k \in \mathcal{K}$ in the system, we define the right-hand-side of (\ref{VFuncn}) as a Q-factor, which is a mapping $Q_k: \widehat{\mathcal{S}}_k \times \{0, 1\} \times \mathcal{X} \times$ $\mathcal{D} \rightarrow \mathds{R}$\footnote{To keep what follows uniform, we do not exclude the infeasible decision-makings under a local state for an MU.}, namely,
\begin{align}\label{QFact}
     Q_k\!\left(\widehat{\mathbf{S}}_k, (\varphi_k, X_k, R_k)\right)
 & = (1 - \gamma) \cdot  \ell_k\!\left(\mathbf{S}, (\varphi_k, X_k, R_k)\right) \nonumber\\
 & + \gamma \cdot \sum_{\widehat{\mathbf{S}}_k' \in \widehat{\mathcal{S}}_k}
   \mathbb{P}\!\left(\widehat{\mathbf{S}}_k' | \widehat{\mathbf{S}}_k,
   (\varphi_k, X_k, R_k)\right) \cdot V_k\!\left(\widehat{\mathbf{S}}_k'\right),
\end{align}
where $\varphi_k$, $X_k$ and $R_k$ correspond to, respectively, the channel allocation, the computation task offloading and the input data packet scheduling decisions under the current local state $\widehat{\mathbf{S}}_k$.
For notational simplicity, the channel allocation function $\varphi_k(\bm\beta_k, \bm\pi_{(\mathrm{c}), -k}^*(\widehat{\mathbf{S}}_{- k}))$ of the auction bidding variable $\bm\beta_k$ is equivalently substituted by $\varphi_k$.
By strictly following (\ref{locaPartStatTran}) and (\ref{QFact}), we further define a post-decision Q-factor by
\begin{align}\label{postDeci0}
   \widetilde{Q}_k\!\left(\widetilde{\mathbf{S}}_k, (\varphi_k, X_k, R_k)\right) =
   \gamma \cdot \sum_{\widehat{\mathbf{S}}_k' \in \widehat{\mathcal{S}}_k}
   \mathbb{P}\!\left(\widehat{\mathbf{S}}_k' | \widetilde{\mathbf{S}}_k, (\varphi_k, X_k, R_k)\right) \cdot
   V_k\!\left(\widehat{\mathbf{S}}_k'\right),
\end{align}
which indicates another mapping for MU $k$, that is, $\widetilde{Q}_k: \widehat{\mathcal{S}}_k \times \{0, 1\} \times \mathcal{X} \times \mathcal{D} \rightarrow \mathds{R}$.

By substituting (\ref{postDeci0}) back into (\ref{auctBid0}), we eventually arrive at the true valuation of each MU $k \in \mathcal{K}$,
\begin{align}\label{auctBidNew}
     \nu_k
   = u_k\!\left(\mathbf{S}^j, \left(z_k, \pi_{(\mathrm{t}), k}^*\!\left(\widehat{\mathbf{S}}_k\right),
     \pi_{(\mathrm{p}), k}^*(\mathbf{S}_k)\right)\right)
   + \frac{1}{1 - \gamma} \cdot \widetilde{Q}_k\!\left(\widetilde{\mathbf{S}}_k,
     \left(z_k, \pi_{(\mathrm{t}), k}^*\!\left(\widehat{\mathbf{S}}_k\right), \pi_{(\mathrm{p}), k}^*(\mathbf{S}_k)\right)\right),
\end{align}
where the preference $z_k$ can be then derived from
\begin{align}
     z_k
 & = \underset{z \in \{0, 1\}}{\arg\max}~Q_k\!\left(\widehat{\mathbf{S}}_k,
     \left(z, \pi_{(\mathrm{t}), k}^*\!\left(\widehat{\mathbf{S}}_k\right), \pi_{(\mathrm{p}), k}^*(\mathbf{S}_k)\right)\right),
\end{align}
instead of originally from (\ref{auctBid2}).
In the following subsection, we propose a novel deep RL scheme to learn the Q-factor and the post-decision Q-factor for each MU $k$.

\subsection{Proposed Deep RL Scheme}

With the previously defined Q-factor as in (\ref{QFact}), the optimal state-value function for each MU $k \in \mathcal{K}$ in the system can be in turn obtained from
\begin{align}\label{StatValu1}
  V_k\!\left(\widehat{\mathbf{S}}_k\right) =
  \max_{\varphi_k, X_k, R_k} Q_k\!\left(\widehat{\mathbf{S}}_k, (\varphi_k, X_k, R_k)\right),
\end{align}
$\forall \widehat{\mathbf{S}}_k \in \widehat{\mathcal{S}}_k$.
The conventional model-free Q-learning algorithm can be applied to learn both the Q-factor and the post-decision Q-factor \cite{He19}.
During the learning process, MU $k$ first acquires $\widehat{\mathbf{S}}_k = \widehat{\mathbf{S}}_k^j$, $(\varphi_k, X_k, R_k) = (\varphi_k^j, X_k^j, R_k^j)$, $\ell_k(\mathbf{S}, (\varphi_k, X_k, R_k))$ during a current decision epoch $j$ as well as $\widehat{\mathbf{S}}_k' = \widehat{\mathbf{S}}_k^{j + 1}$ at the beginning of next decision epoch $j + 1$, and then proceeds to update the Q-factor and the post-decision Q-factor in an iterative manner using, respectively,
\begin{align}\label{QFactLear}
 & Q_k^{j + 1}\!\left(\widehat{\mathbf{S}}_k, (\varphi_k, X_k, R_k)\right) =
   Q_k^j\!\left(\widehat{\mathbf{S}}_k, (\varphi_k, X_k, R_k)\right) +                                                  \\
 & \alpha^j \cdot \left((1 - \gamma) \cdot \ell_k\!\left(\mathbf{S}, (\varphi_k, X_k, R_k)\right) +
   \gamma \cdot \max_{\varphi_k', X_k', R_k'} Q_k^j\!\left(\widehat{\mathbf{S}}_k', (\varphi_k', X_k', R_k')\right) -
   Q_k^j\!\left(\widehat{\mathbf{S}}_k, (\varphi_k, X_k, R_k)\right)\right),                                            \nonumber
\end{align}
and
\begin{align}\label{postDeciLear}
     \widetilde{Q}_k^{j + 1}\!\left(\widetilde{\mathbf{S}}_k, (\varphi_k, X_k, R_k)\right)
 & = \widetilde{Q}_k^j\!\left(\widetilde{\mathbf{S}}_k, (\varphi_k, X_k, R_k)\right)                        \\
 & + \alpha^j \cdot \left(\gamma \cdot \max_{\varphi_k', X_k', R_k'} Q_k^j\!\left(\widehat{\mathbf{S}}_k',
     (\varphi_k', X_k', R_k')\right) -
     \widetilde{Q}_k^j\!\left(\widetilde{\mathbf{S}}_k, (\varphi_k, X_k, R_k)\right)\right),                \nonumber
\end{align}
where $\alpha^j \in [0, 1)$ denotes the learning rate.
It has been well established that if: 1) the global system state transition probability under $\bm\pi^*$ is time-invariant; 2) $\sum_{j = 1}^\infty \alpha^j$ is infinite and $\sum_{j = 1}^\infty (\alpha^j)^2$ is finite; and 3) the finite space $\widehat{\mathcal{S}}_k \times$ $\{0, 1\} \times \mathcal{X} \times \mathcal{D}$ is exhaustively explored, the learning process surely converges \cite{He19, Mast13}.

\begin{figure*}[!t]
  \centering
  \includegraphics[width=28pc]{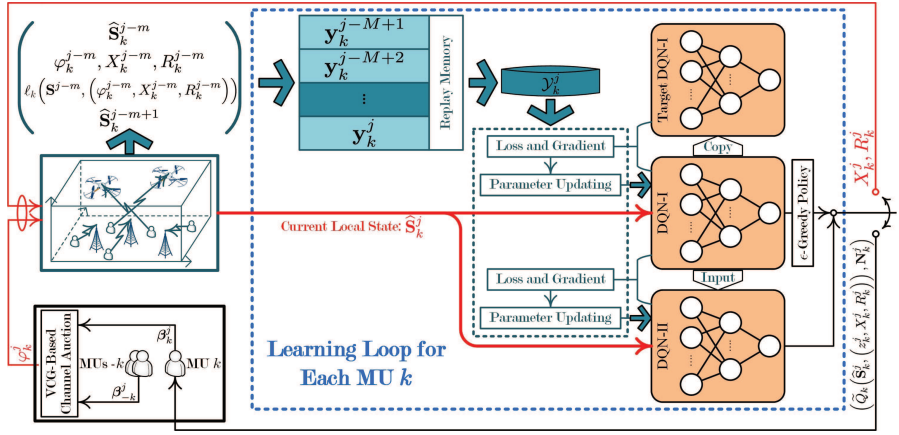}
  \caption{Implementation of the proposed deep reinforcement learning scheme to approach the Q-factor and the post-decision Q-factor of each mobile user (MU) $k \in \mathcal{K}$ in the system (DQN: deep Q-network.).}
  \label{deepLear}
\end{figure*}

It is not difficult to find that for the air-ground integrated MEC system investigated in this paper, the space $\widehat{\mathcal{S}}_k$ of local states faced by each MU $k \in \mathcal{K}$ is extremely huge.
The tabular nature in representing the Q-factor and the post-decision Q-factor values makes the learning rule as in (\ref{QFactLear}) and (\ref{postDeciLear}) impractical.
Inspired by the recent advances in neural networks \cite{Appl17} and the widespread success of a deep neural network \cite{Mnih15}, we propose to adopt two separate deep Q-networks (DQNs), namely, DQN-I and DQN-II, to reproduce the Q-factor and the post-decision Q-factor of an MU.
More specifically, for each MU $k$, we model the Q-factor in (\ref{QFact}) by
\begin{align}
        Q_k\!\left(\widehat{\mathbf{S}}_k, (\varphi_k, X_k, R_k)\right)
\approx Q_k\!\left(\widehat{\mathbf{S}}_k, (\varphi_k, X_k, R_k); \bm\theta_k\right),
\end{align}
$\forall (\widehat{\mathbf{S}}_k, (\varphi_k, X_k, R_k)) \in \widehat{\mathcal{S}}_k \times \{0, 1\} \times \mathcal{X} \times \mathcal{D}$, and the post-decision Q-factor in (\ref{postDeci0}) by
\begin{align}
        \widetilde{Q}_k\!\left(\widetilde{\mathbf{S}}_k, (\varphi_k, X_k, R_k)\right)
\approx \widetilde{Q}_k\!\left(\widetilde{\mathbf{S}}_k, (\varphi_k, X_k, R_k); \tilde{\bm\theta}_k\right),
\end{align}
$\forall (\widetilde{\mathbf{S}}_k, (\varphi_k, X_k, R_k))$ $\in \widehat{\mathcal{S}}_k \times \{0, 1\} \times \mathcal{X} \times \mathcal{D}$, where $\bm\theta_k$ and $\tilde{\bm\theta}_k$ denote, respectively, the vectors of parameters that are associated with DQN-I of the Q-factor and DQN-II of the post-decision Q-factor.
Similar to the A2C architecture \cite{Mnih16}, DQN-I with $\bm\theta_k$ of MU $k$ in the proposed deep RL scheme estimates the Q-factor values while DQN-II with $\tilde{\bm\theta}_k$ approximates the best-response control policy suggested by DQN-I \cite{Ram10, Wang19}.
MU $k$ learns $\bm\theta_k$ and $\tilde{\bm\theta}_k$, rather than finding the Q-factor and the post-decision Q-factor values according to (\ref{QFactLear}) and (\ref{postDeciLear}).
The implementation of the proposed deep RL scheme is illustrated in Fig. \ref{deepLear}.

During the deep RL process, each MU $k \in \mathcal{K}$ in the system is equipped with a finite replay memory $\mathcal{M}_k^j = \{\mathbf{y}_k^{j - M + 1},$ $\cdots, \mathbf{y}_k^j\}$ to store the most recent $M$ historical experiences up to a decision epoch $j$, where an experience $\mathbf{y}_k^{j - m + 1}$ ($1 \leq m \leq M$) given by
\begin{align}\label{exper}
       \mathbf{y}_k^{j - m + 1}
 = \left(\widehat{\mathbf{S}}_k^{j - m}, \left(\varphi_k^{j - m}, X_k^{j - m}, R_k^{j - m}\right),
   \ell_k\!\left(\mathbf{S}^{j - m}, \left(\varphi_k^{j - m}, X_k^{j - m}, R_k^{j - m}\right)\right),
       \widehat{\mathbf{S}}_k^{j - m + 1}\right),
\end{align}
happens at the transition between two consecutive decision epochs $j - m$ and $j - m + 1$.

\subsubsection{DQN-I Training}
Each MU $k \in \mathcal{K}$ maintains an DQN-I as well as a target DQN-I, which are $Q_k(\widehat{\mathbf{S}}_k, (\varphi_k, X_k, R_k); \bm\theta_k^j)$ and $Q_k(\widehat{\mathbf{S}}_k, (\varphi_k, X_k, R_k); \bm\theta_k^{j, -})$ with $\bm\theta_k^j$ and $\bm\theta_k^{j, -}$ being the associated vectors of parameters at each decision epoch $j$ and from a previous decision epoch before epoch $j$, respectively.
To perform experience replay \cite{Lin92}, MU $k$ randomly samples a mini-batch $\mathcal{Y}_k^j \subseteq \mathcal{M}_k^j$ from the replay memory $\mathcal{M}_k^j$ at each decision epoch $j$ to train DQN-I.
The training objective is to update the parameters $\bm\theta_k^j$ of DQN-I in the direction of minimizing the loss function $\textsf{LOSS}_{(\mbox{\footnotesize DQN-I}), k}(\bm\theta_k^j)$,
\begin{align}\label{lossFunc1}
&   \textsf{LOSS}_{(\mbox{\footnotesize DQN-I}), k}\!\left(\bm\theta_k^j\right)
  = \textsf{E}_{\left\{\left(\widehat{\mathbf{S}}_k, \left(\varphi_k, X_k, R_k\right),
    \ell_k\!\left(\mathbf{S}, \left(\varphi_k, X_k, R_k\right)\right), \widehat{\mathbf{S}}_k'\right)
    \in \mathcal{Y}_k^j\right\}}\!\!
    \left[\left((1 - \gamma) \cdot
    \ell_k\!\left(\mathbf{S}, \left(\varphi_k, X_k, R_k\right)\right)
    \vphantom{\underset{\varphi_k', X_k', R_k'}{\arg\max}}\right.
    \vphantom{\left(\underset{\varphi_k', X_k', R_k'}{\arg\max}\right)^2}\right. +                                   \nonumber\\
& \left.\left.\gamma \cdot Q_k\!\left(\widehat{\mathbf{S}}_k',
                                  \underset{\varphi_k', X_k', R_k'}{\arg\max}
                                  Q_k\!\left(\widehat{\mathbf{S}}_k', \left(\varphi_k', X_k', R_k'\right); \bm\theta_k^j\right); \bm\theta_k^{j, -}\right) -
   Q_k\!\left(\widehat{\mathbf{S}}_k, \left(\varphi_k, X_k, R_k\right); \bm\theta_k^j\right)\right)^2\right].
\end{align}
By differentiating $\textsf{LOSS}_{(\mbox{\footnotesize DQN-I}), k}(\bm\theta_k^j)$ with respect to $\bm\theta_k^j$, we obtain the gradient as
\begin{align}\label{grad1}
 &   \nabla_{\bm\theta_k^j} \textsf{LOSS}_{(\mbox{\footnotesize DQN-I}), k}\!\left(\bm\theta_k^j\right) =
     \textsf{E}_{\left\{\left(\widehat{\mathbf{S}}_k, \left(\varphi_k, X_k, R_k\right),
     \ell_k\!\left(\mathbf{S}, \left(\varphi_k, X_k, R_k\right)\right), \widehat{\mathbf{S}}_k'\right)
     \in \mathcal{Y}_k^j\right\}}\!\!
     \left[\left((1 - \gamma) \cdot
     \ell_k\!\left(\mathbf{S}, \left(\varphi_k, X_k, R_k\right)\right) +
     \vphantom{\underset{\varphi_k', X_k', R_k'}{\arg\max}}\right.\right.                                          \nonumber\\
 &   \left.\left.\gamma \cdot Q_k\!\left(\widehat{\mathbf{S}}_k',
                                  \underset{\varphi_k', X_k', R_k'}{\arg\max}
                                  Q_k\!\left(\widehat{\mathbf{S}}_k', \left(\varphi_k', X_k', R_k'\right); \bm\theta_k^j\right); \bm\theta_k^{j, -}\right) -
    Q_k\!\left(\widehat{\mathbf{S}}_k, \left(\varphi_k, X_k, R_k\right); \bm\theta_k^j\right)\right)\right. \cdot\nonumber\\
 &  \left.\nabla_{\bm\theta_k^j} Q_k\!\left(\widehat{\mathbf{S}}_k, \left(\varphi_k, X_k, R_k\right); \bm\theta_k^j\right)
    \vphantom{\underset{\varphi_k', X_k', R_k'}{\arg\max}}\right].
\end{align}

\subsubsection{DQN-II Training}
At each decision epoch $j$, we designate $\tilde{\bm\theta}_k^j$ as the parameters associated with DQN-II of each MU $k \in \mathcal{K}$ in the system.
Taking $\bm\theta_k^j$ from DQN-I as an input, MU $k$ updates $\tilde{\bm\theta}_k^j$ to minimize the loss function $\textsf{LOSS}_{(\mbox{\footnotesize DQN-II}), k}(\tilde{\bm\theta}_k^j)$ given by
\begin{align}\label{lossFunc2}
    \textsf{LOSS}_{(\mbox{\footnotesize DQN-II}), k}\!\left(\tilde{\bm\theta}_k^j\right)
& = \textsf{E}_{\left\{\left(\widehat{\mathbf{S}}_k, \left(\varphi_k, X_k, R_k\right),
    \ell_k\!\left(\mathbf{S}, \left(\varphi_k, X_k, R_k\right)\right), \widehat{\mathbf{S}}_k'\right)
    \in \mathcal{Y}_k^j\right\}}\!\!
    \Big[\Big(\gamma \cdot \max\limits_{\varphi_k', X_k', R_k'}
    Q_k\!\left(\widehat{\mathbf{S}}_k', \left(\varphi_k', X_k', R_k'\right); \bm\theta_k^j\right) \nonumber\\
& - \widetilde{Q}_k\!\left(\widetilde{\mathbf{S}}_k, \left(\varphi_k, X_k, R_k\right); \tilde{\bm\theta}_k^j\right)\Big)^2\Big],
\end{align}
over the mini-batch $\mathcal{Y}_k^j$ using the gradient as
\begin{align}\label{grad2}
     \nabla_{\tilde{\bm\theta}_k^j} \textsf{LOSS}_{(\mbox{\footnotesize DQN-II}), k}\!\left(\tilde{\bm\theta}_k^j\right)
 & = \textsf{E}_{\left\{\left(\widehat{\mathbf{S}}_k, \left(\varphi_k, X_k, R_k\right),
     \ell_k\!\left(\mathbf{S}, \left(\varphi_k, X_k, R_k\right)\right), \widehat{\mathbf{S}}_k'\right)
     \in \mathcal{Y}_k^j\right\}}\!\!
     \left[\left(\gamma \cdot \max\limits_{\varphi_k', X_k', R_k'}
     Q_k\!\left(\widehat{\mathbf{S}}_k', \left(\varphi_k', X_k', R_k'\right); \bm\theta_k^j\right)\right.\right. \nonumber\\
 & - \left.\left.\widetilde{Q}_k\!\left(\widetilde{\mathbf{S}}_k, \left(\varphi_k, X_k, R_k\right);
     \tilde{\bm\theta}_k^j\right)\vphantom{\max\limits_{\varphi_k', X_k', R_k'}}\!\right) \cdot \nabla_{\tilde{\bm\theta}_k^j}
     \widetilde{Q}_k\!\left(\widetilde{\mathbf{S}}_k, \left(\varphi_k, X_k, R_k\right); \tilde{\bm\theta}_k^j\right)\right].
\end{align}

In Algorithm \ref{algo}, we briefly summarize the procedure of the proposed online deep RL scheme implemented by each MU $k \in \mathcal{K}$ in the air-ground integrated MEC system.

\begin{algorithm}[!t]
    \caption{Online Deep RL Scheme for Learning Q-Factor and Post-Decision Q-Factor of Each MU $k \in \mathcal{K}$}
    \label{algo}
    \begin{algorithmic}[1]
        \STATE \textbf{initialize} the replay memory $\mathcal{M}_k^j$ of size $M \in \mathds{N}_+$, the mini-batch $\mathcal{Y}_k^j$, an DQN-I, a target DQN-I and an DQN-II with parameters $\bm\theta_k^j$, $\bm\theta_k^{j, -}$ and $\tilde{\bm\theta}_k^j$, and the local state $\widehat{\mathbf{S}}_k^j$, for the initial decision epoch $j = 1$.

        \REPEAT
            \STATE At the beginning of decision epoch $j$, MU $k$ first takes the observation of $\widehat{\mathbf{S}}_k^j$ as an input to DQN-I with parameters $\bm\theta_k^j$, and then selects $(z_k^j, X_k^j, R_k^j)$ randomly with probability $\epsilon$ or $(z_k^j, X_k^j, R_k^j)$ that is with maximum value $Q_k(\widehat{\mathbf{S}}_k^j, (z_k^j, X_k^j, R_k^j); \bm\theta_k^j)$ with probability $1 - \epsilon$.

            \STATE MU $k$ computes the auction bid $\bm\beta_k^j = (\nu_k^j, \mathbf{N}_k^j)$ according to (\ref{auctBidNew}), (\ref{auctBid1_0}) and (\ref{auctBid1_1}) with $\widetilde{Q}_k(\widetilde{\mathbf{S}}_k^j, (\varphi_k, X_k, R_k)) \approx \widetilde{Q}_k(\widetilde{\mathbf{S}}_k^j, (\varphi_k, X_k, R_k); \tilde{\bm\theta}_k^j)$.

            \STATE MU $k$ sends $\bm\beta_k^j$ to the RO of the third-party real-time application SP.

            \STATE With the bids from all MUs, the RO determines the auction winners $\bm\phi^j$ and the channel allocation $\bm\rho_k^j$ according to (\ref{chanSche}), and calculates the payments $\tau_k^j$ according to (\ref{paymCalc}).

            \STATE With the channel allocation $\bm\rho_k^j$, MU $k$ makes computation offloading $X_k^j$ and packet scheduling $\varphi_k^j \cdot R_k^j$.

            \STATE MU $k$ achieves the payoff value $\ell_k(\mathbf{S}_k^j, (\varphi_k, X_k, R_k))$ and observes $\widehat{\mathbf{S}}_k^{j + 1}$ at the next decision epoch $j + 1$.

            \STATE MU $n$ updates the replay memory $\mathcal{M}_k^j$ with $\mathbf{m}_k^j$.

            \STATE With a randomly sampled $\mathcal{Y}_k^j$ from $\mathcal{M}_k^j$, MU $k$ updates $\bm\theta_k^j$ of DQN-I and $\tilde{\bm\theta}_k^j$ of DQN-II with the gradients in (\ref{grad1}) and (\ref{grad2}), respectively.

            \STATE MU $k$ regularly resets the target DQN-I parameters with $\bm\theta_k^{j + 1, -} = \bm\theta_k^j$, and otherwise $\bm\theta_k^{j + 1, -} = \bm\theta_k^{j, -}$.

            \STATE The decision epoch index is updated by $j \leftarrow j + 1$.
        \UNTIL{A predefined stopping condition is satisfied.}
    \end{algorithmic}
\end{algorithm}

\section{Numerical Experiments}
\label{simu}

In order to quantitatively evaluate the performance gained from the proposed deep RL scheme, we conduct numerical experiments based on TensorFlow \cite{Abad16}.

\subsection{Parameter Settings}
We set up an experimental scenario of the RAN covering a $0.4\times0.4$ Km$^2$ square area, where there are $B = 4$ BSs and $|\mathcal{K}| = 20$ MUs.
The BSs are placed at equal distance apart, and the square area is divided into $|\mathcal{L}| = 1600$ locations with each representing a small area of $10\times10$ m$^2$.
The flying altitude of the UAV is kept to $H = 100$ meters.
For each MU $k \in \mathcal{K}$ in the system, $G_{b, k}^j$ and $G_{(\mathrm{\mathrm{v}}), k}^j$, $\forall b \in \mathcal{B}$ and $\forall j$, follow the channel model in \cite{Chen19} and the LOS model in \cite{Zeng16}, respectively.
The state transition probability matrices underlying the Markov mobilities of the UAV and all MUs are independently and randomly generated.
We design the DQN-I and the DQN-II of an MU to be with two hidden layers, each of which contains $32$ neurons.
ReLU is selected as the activation function \cite{Nair10} and Adam as the optimizer \cite{King15}.
Other parameter values are listed in Table \ref{tabl1}.
\begin{table}[t]
  \caption{Parameter values in experiments.}\label{tabl1}
        \begin{center}
        \begin{tabular}{|c|c||c|c|}
              \hline
              Parameter       & Value                     & Parameter     & Value                           \\\hline
              \hline
              $D_{(\max)}$    & $10$                      & $\mu$         & $500$ Kbits                     \\\hline
              $\vartheta$     & $1300$                    & $A_{(\max)}$  & $30$ seconds                    \\\hline
              $\eta$          & $1$ MHz                   & $\sigma^2$    & $-144$ dBm/Hz                   \\\hline
              $\delta$        & $1$ second                & $P_{(\max)}$  & $3$ Watt                        \\\hline
              $\varpi_k$      & $10$, $\forall k$         & $\omega_k$    & $2$, $\forall k$                \\\hline
              $\varrho$       & $1$ GHz                   & $\xi$         & $10^{-2}$ seconds               \\\hline
              $\chi_0$        & $2\cdot 10^7$ bits/second & $\varepsilon$ & $0.2$                           \\\hline
              $\varsigma$     & $10^{-27}$                & $M$           & $5000$                          \\
              \hline
        \end{tabular}
        \end{center}
\end{table}

For the performance comparisons, we develop the following four baseline schemes as well.
\begin{enumerate}
  \item \emph{Local Computation (Baseline 1)} -- Each MU processes the arriving computation tasks only at the local mobile device, and hence no channel auction is involved.
  \item \emph{Server Execution (Baseline 2)} -- Each MU always offloads the computations to the ground MEC server for execution.
  \item \emph{UAV Execution (Baseline 3)} -- All computation tasks from the pre-processing buffer of each MU are processed by the VMs at the UAV.
  \item \emph{Greedy Processing (Baseline 4)} -- Whenever possible, a buffered computation task is computed locally or executed remotely via the better link of the two between the MU and the server as well as the UAV.
\end{enumerate}
Implementing Baselines 2, 3 and 4 during each decision epoch, an MU defines the valuation of winning the channel auction as the utility that can be potentially achieved from transmitting a maximum number of input data packets.

\subsection{Experiment Results}

\subsubsection{Experiment 1 -- Convergence Performance}

\begin{figure}[t]
  \centering
  \includegraphics[width=0.32\textwidth]{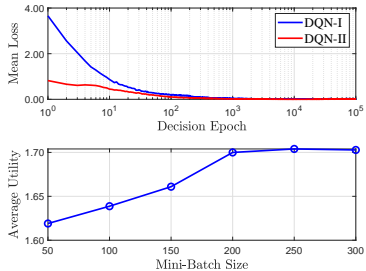}
  \caption{Illustration of convergence speed of the proposed deep RL scheme in terms of mean losses (namely, $(1 / |\mathcal{K}|) \cdot \sum_{k \in \mathcal{K}} \textsf{LOSS}_{(\mbox{\footnotesize DQN-I}), k}(\bm\theta_k^j)$ and $(1 / |\mathcal{K}|) \cdot \sum_{k \in \mathcal{K}} \textsf{LOSS}_{(\mbox{\footnotesize DQN-II}), k}(\tilde{\bm\theta}_k^j)$) versus decision epoch $j$ (upper) and average utility performance per MU across the learning procedure versus batch sizes (lower).}
  \label{simu01}
\end{figure}

The goal of the first experiment is to validate if the air-ground integrated MEC system remains stable when implementing the proposed online deep RL scheme for information freshness-aware task offloading.
We fix the computation task arriving probability and the number of channels to be $\lambda = 0.3$ and $|\mathcal{C}| = 18$, respectively.
For each MU $k \in \mathcal{K}$, we set the mini-batch size as $|\mathcal{Y}_k^j| = 200$, $\forall j$.
We plot the variations in the mean losses $(1 / |\mathcal{K}|) \cdot \sum_{k \in \mathcal{K}} \textsf{LOSS}_{(\mbox{\footnotesize DQN-I}), k}(\bm\theta_k^j)$ and $(1 / |\mathcal{K}|) \cdot \sum_{k \in \mathcal{K}} \textsf{LOSS}_{(\mbox{\footnotesize DQN-II}), k}(\tilde{\bm\theta}_k^j)$ over all the MUs versus the decision epochs in the upper subplot in Fig. \ref{simu01}, which shows that the proposed scheme converges within $10^4$ epochs.
In the lower subplot in Fig. \ref{simu01}, we plot the average utility performance per MU with various mini-batch sizes under the given replay memory capacity.
It is obvious from (\ref{grad1}) and (\ref{grad2}) that for each MU, a larger mini-batch size results in a more stable gradient estimate, i.e., a smaller variance, hence a better average utility performance across the learning procedure.
When the mini-batch size exceeds $200$, the average utility performance improvement saturates.
In Experiments 2 and 3, we hence continue to use a mini-batch of size $200$ for all MUs to strike a balance between the performance improvement and the computational overhead.

\subsubsection{Experiment 2 -- Performance under Different Task Arriving Probabilities}

\begin{figure}[!htb]
\minipage{0.32\textwidth}
    \centering
    \includegraphics[width=\linewidth]{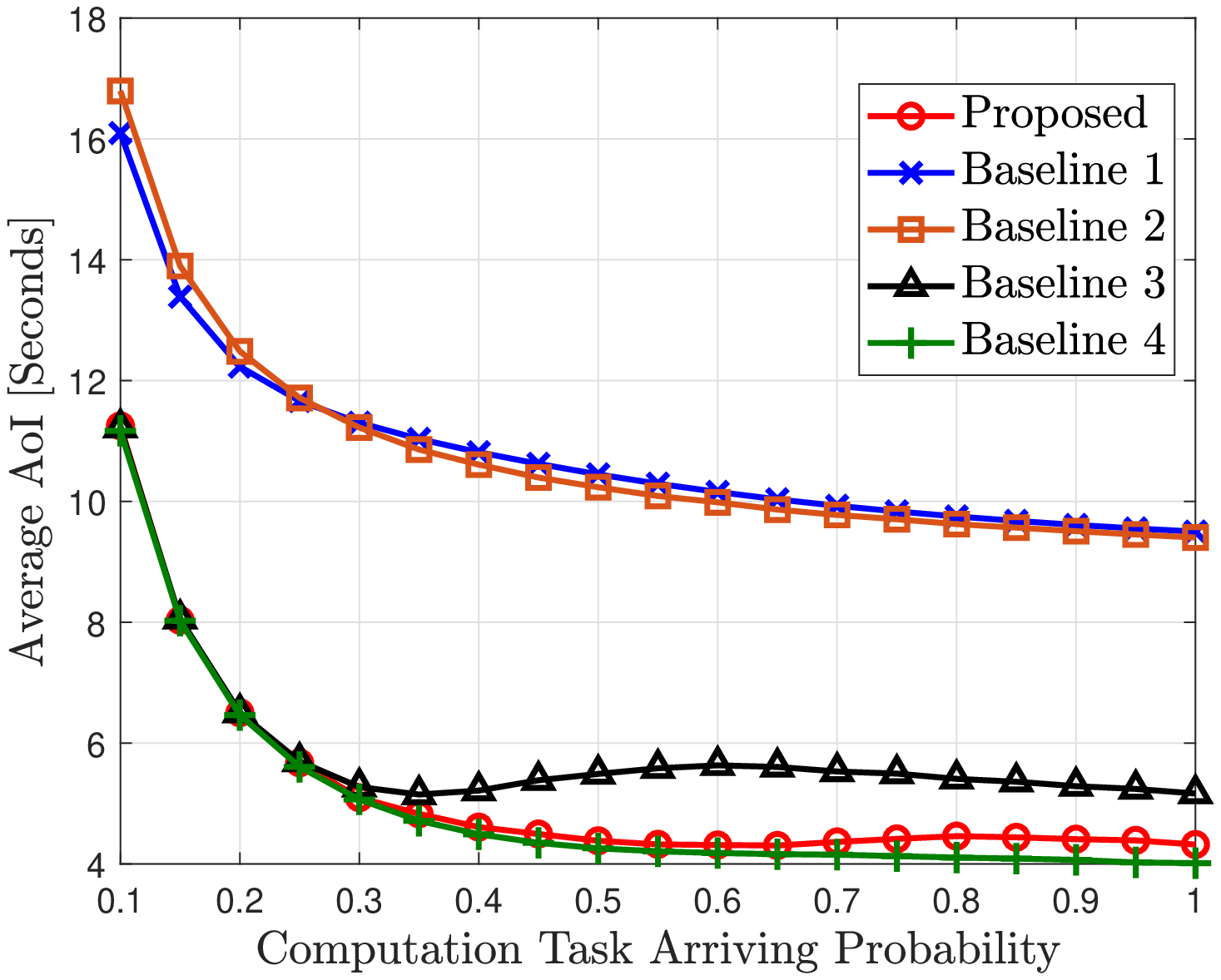}
    \caption{Average AoI performance per MU across the learning procedure versus computation task arriving probability.}
    \label{simu02_01}
\endminipage\hfill
\minipage{0.32\textwidth}
    \centering
    \includegraphics[width=\linewidth]{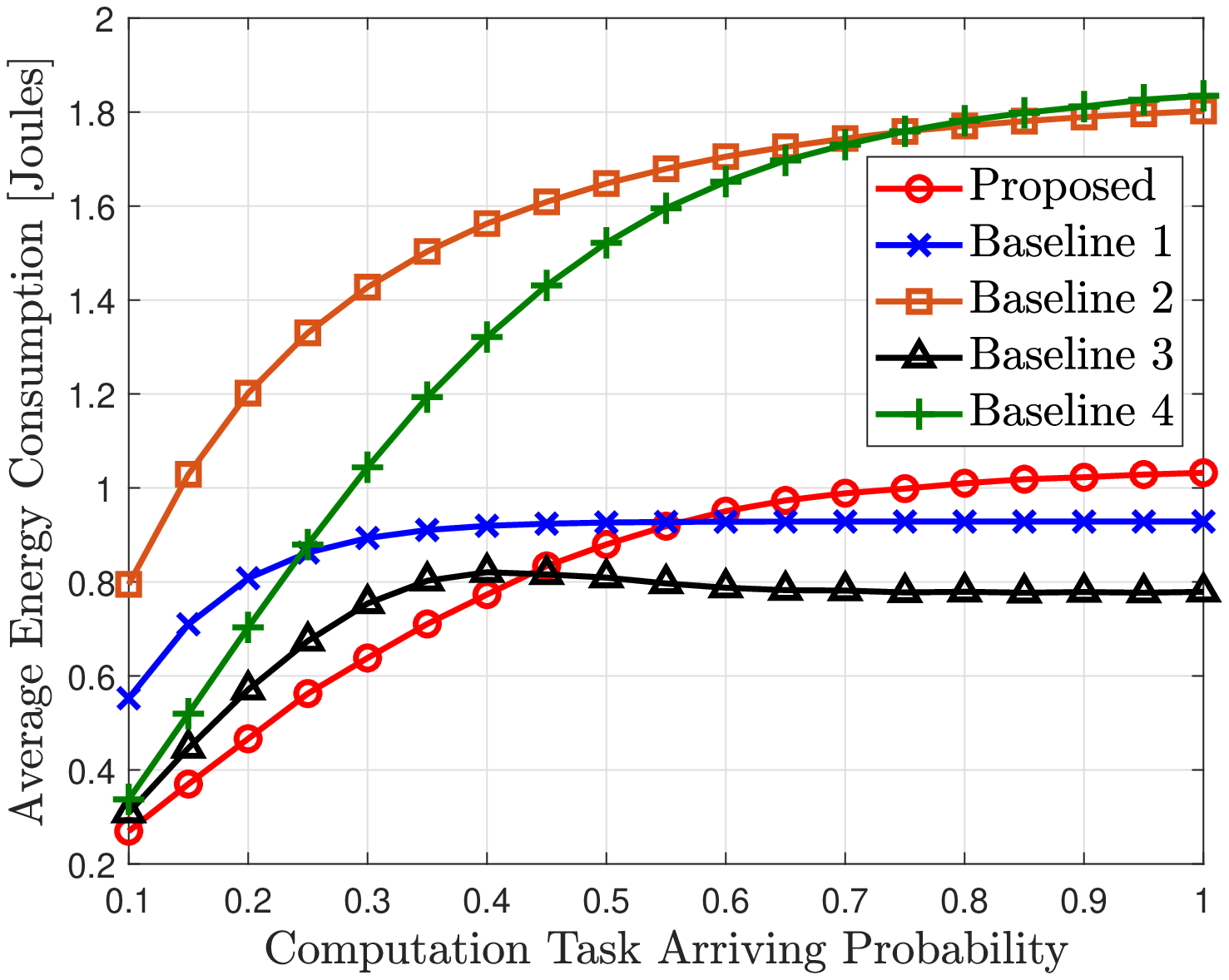}
    \caption{Average energy consumption per MU across the learning procedure versus computation task arriving probability.}
    \label{simu02_02}
\endminipage\hfill
\minipage{0.32\textwidth}%
    \centering
    \includegraphics[width=\linewidth]{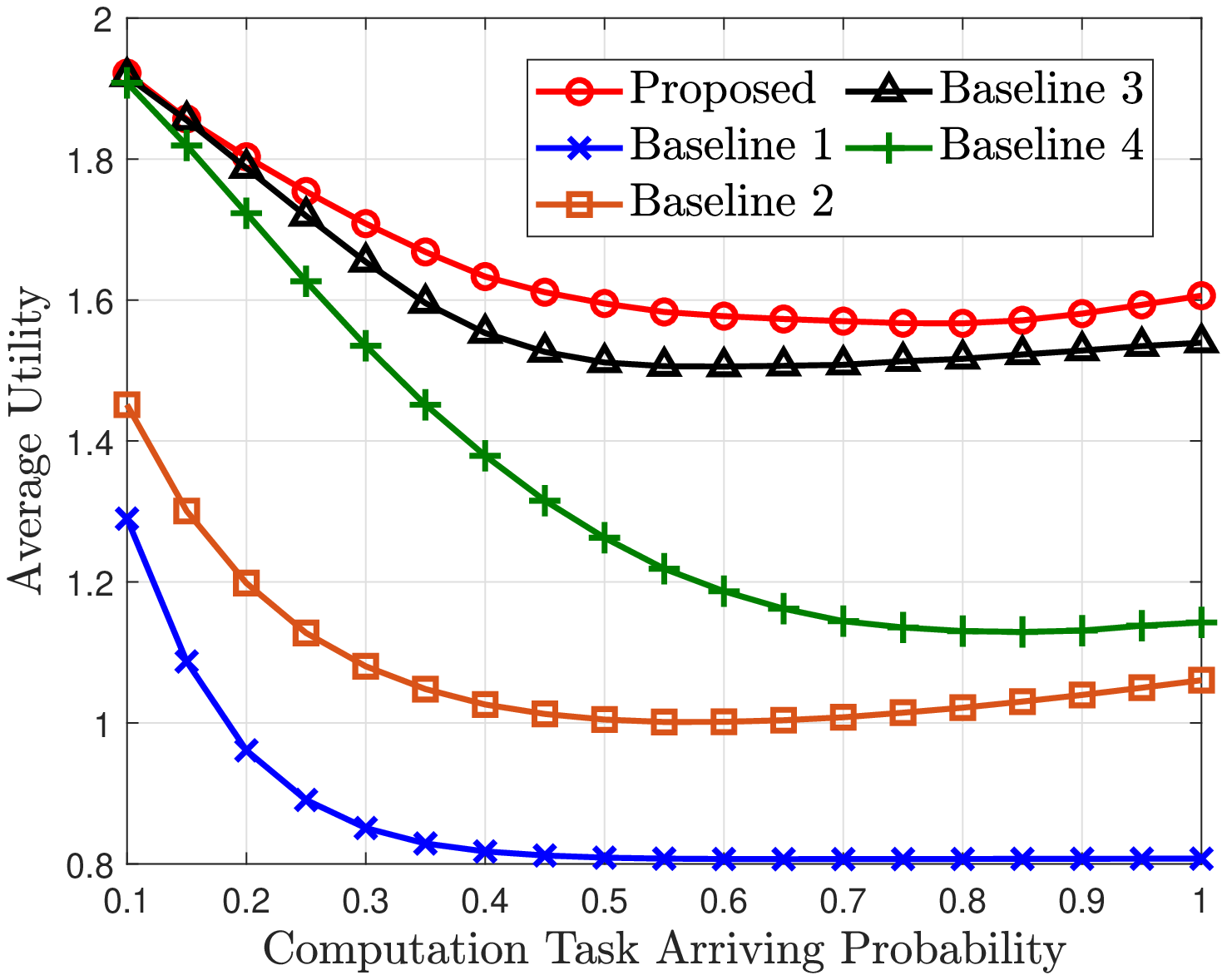}
    \caption{Average utility performance per MU across the learning procedure versus computation task arriving probability.}
    \label{simu02_03}
\endminipage
\end{figure}

In this experiment, we aim to demonstrate the average performance per MU per decision epoch in terms of the average AoI, the average energy consumption and the average utility under different computation task arriving probabilities.
We assume there are $|\mathcal{C}| = 16$ channels in the system, which can be utilized among the non-cooperative MUs to access the computing service provided by the third-party real-time application SP.
The simulated results are exhibited in Figs. \ref{simu02_01}, \ref{simu02_02} and \ref{simu02_03}.
Fig. \ref{simu02_01} illustrates the average AoI per MU.
Fig. \ref{simu02_02} illustrates the average energy consumption per MU.
Fig. \ref{simu02_03} illustrates the average utility per MU.

Each plot compares the performance of the proposed deep RL scheme with the four baseline task offloading schemes.
From Fig. \ref{simu02_03}, it can be observed that the proposed scheme achieves the best performance in average utility per MU.
Fig. \ref{simu02_01} shows that the comparable average AoI performance can be realized between the proposed scheme and Baseline 4.
As the computation task arriving probability increases, each MU consumes more energy for task processing in order to maintain the information freshness, as can be seen from Fig. \ref{simu02_02}.
Note that when implementing Baseline 3, the average energy consumption per MU first increases and then decreases, which is due to the fact that the maximum transmit power at the mobile device of each MU and the constrained computation service rate of an VM at the UAV limit the transmissions of input data packets during a decision epoch.
Similar observations can be made from the curves of the proposed scheme and Baseline 3 in Figs. \ref{simu02_01}, \ref{simu03_01} and \ref{simu03_02}.
On the other hand, Baselines 1, 2 and 4 show monotonic performance in the average AoI and the average energy consumption, as can be expected.
With the chosen weighting constant values, the AoI increasingly dominates the utility function value as the energy consumption increases, which conforms the average utility performance trends of the proposed scheme as well as Baselines 2, 3 and 4.

\subsubsection{Experiment 3 -- Performance with Changing Number of Channels}

\begin{figure}[!htb]
\minipage{0.32\textwidth}
    \centering
    \includegraphics[width=\linewidth]{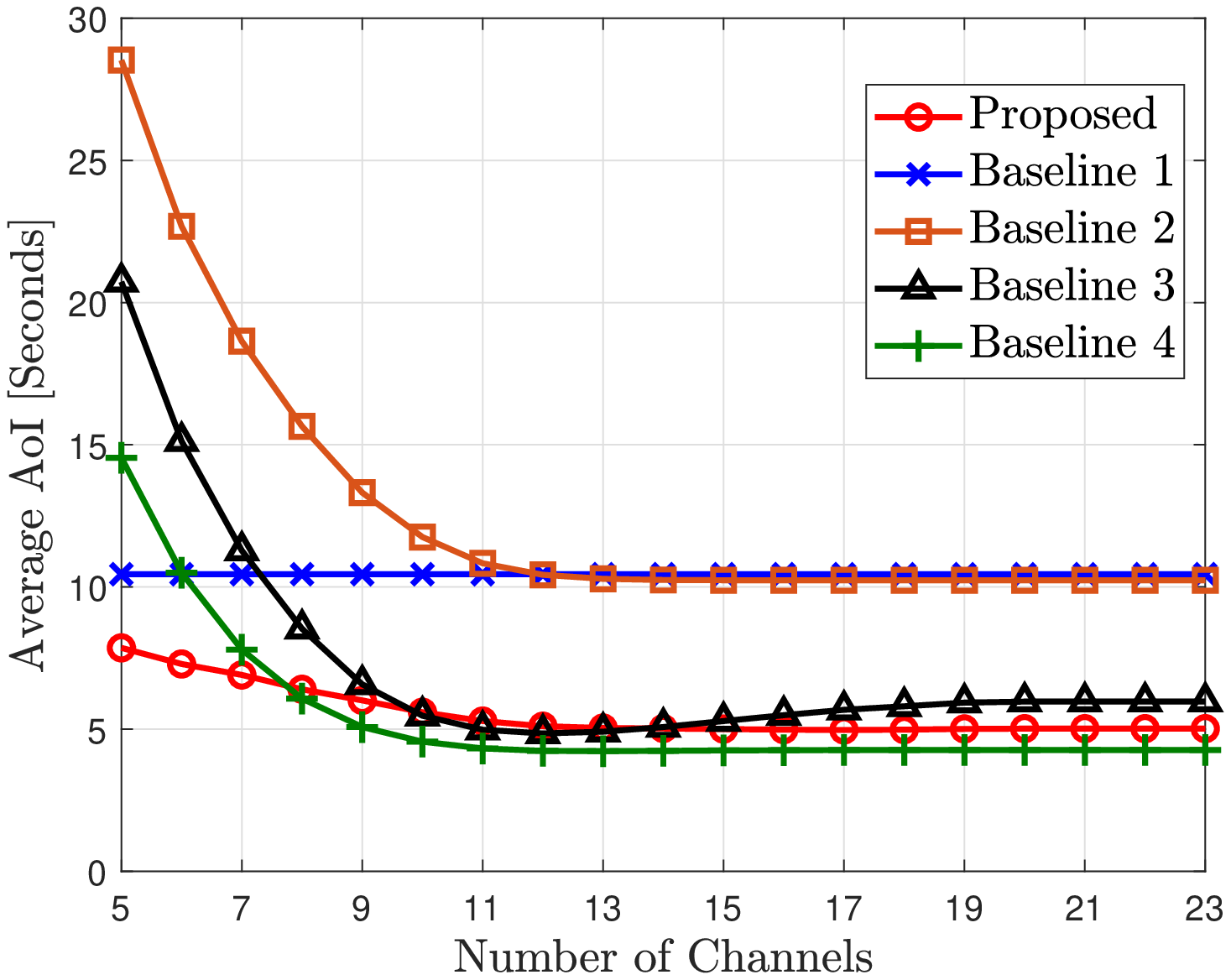}
    \caption{Average AoI performance per MU across the learning procedure versus number of channels.}
    \label{simu03_01}
\endminipage\hfill
\minipage{0.32\textwidth}
    \centering
    \includegraphics[width=\linewidth]{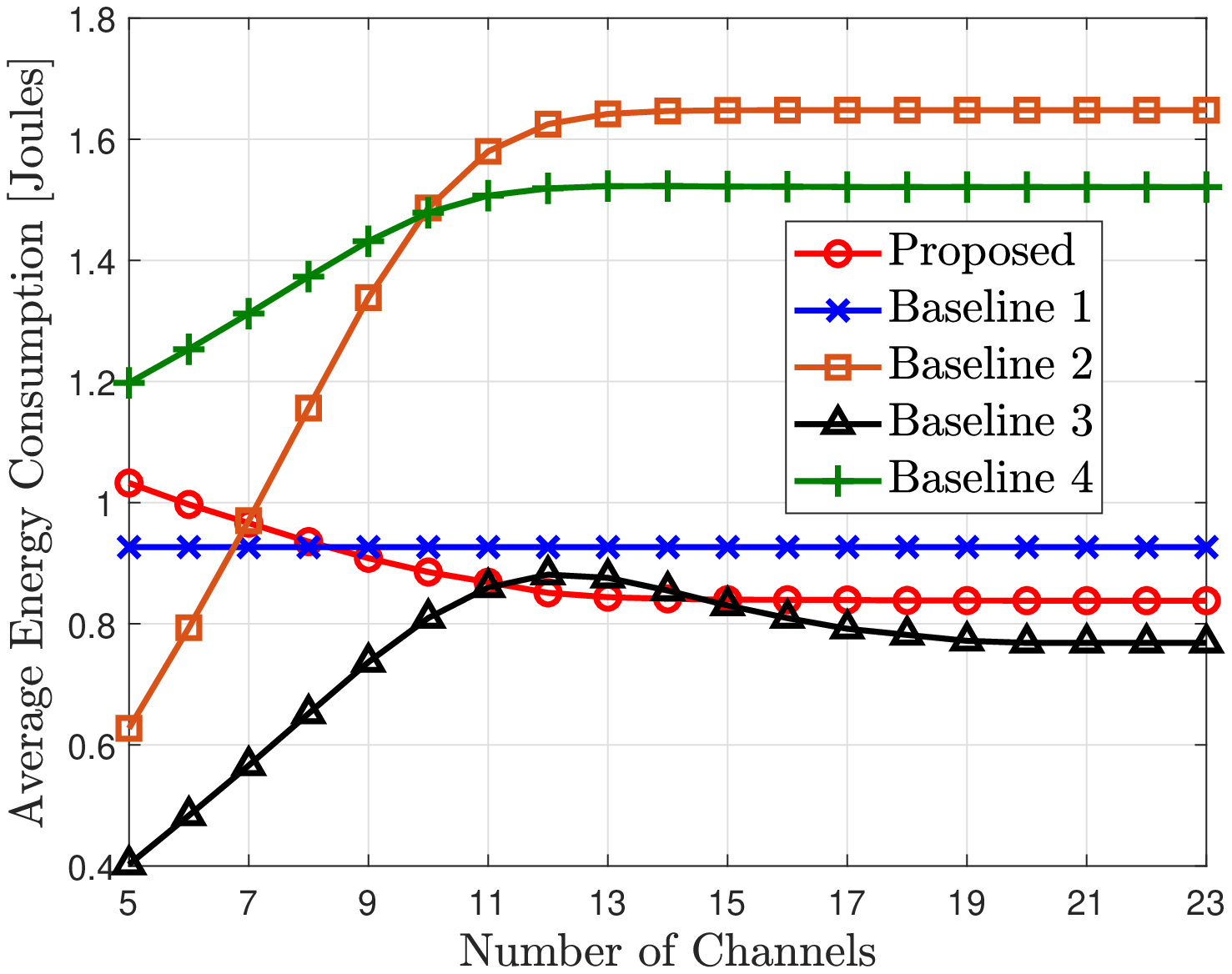}
    \caption{Average energy consumption per MU across the learning procedure versus number of channels.}
    \label{simu03_02}
\endminipage\hfill
\minipage{0.32\textwidth}%
    \centering
    \includegraphics[width=\linewidth]{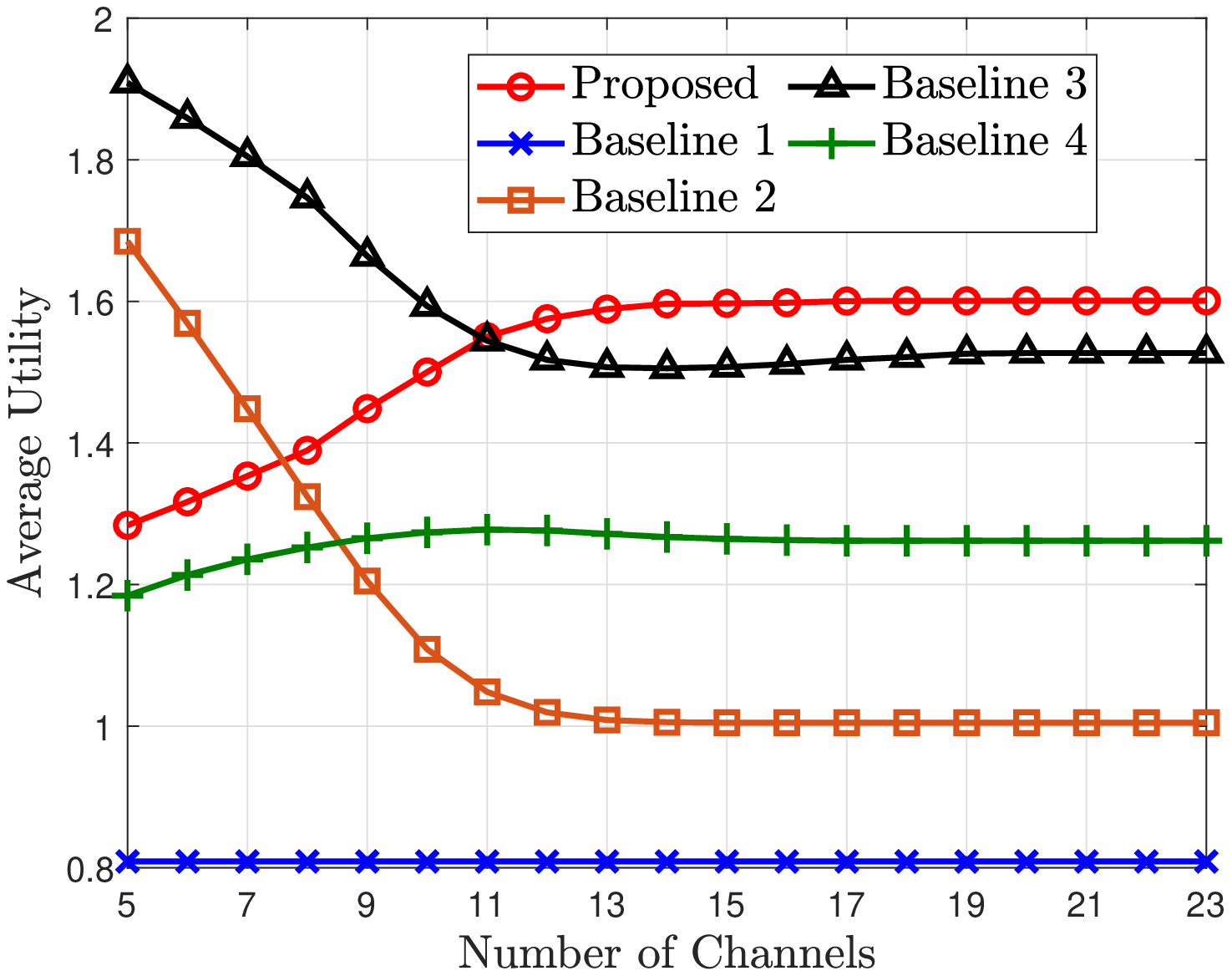}
    \caption{Average utility performance per MU across the learning procedure versus number of channels.}
    \label{simu03_03}
\endminipage
\end{figure}

The last experiment simulates the average performance per MU per decision epoch from the proposed online deep RL scheme and the four baselines versus the numbers of channels.
In experiment, the computation task arriving probability is selected as $\lambda = 0.5$.
The average AoI, average energy consumption and average utility per MU across the entire learning procedure are depicted in Figs. \ref{simu03_01}, \ref{simu03_02} and \ref{simu03_03}, respectively.
It can be easily observed from Fig. \ref{simu03_01} that as the number of available channels increases, the average AoI decreases.
The more channels available in the system, the more likely an MU is able to obtain one channel from the auction.
Therefore, with Baselines 2, 3 and 4, the MU consumes more energy to offload more input data packets for remote execution, while with the proposed deep RL scheme, there are more opportunities for the MU to have a computation task executed remotely with less energy consumption compared with the local processing, as shown in Fig. \ref{simu03_02}.
Though the average AoI from the proposed scheme is smaller than that from Baseline 3, the weight choices in utility function make Baseline 3 outperforming the proposed scheme in average utility when the number of channels is small, as explained in Experiment 2.
Since all MUs do not participate the channel auction, the average performance of Baseline 1 does not change.
Last but not least, both Experiments 2 and 3 tell that the proposed deep RL scheme achieves promising average utility performance while keeping the information fresh for the MUs.

\section{Conclusions}
\label{conc}

In this paper, the purpose is to optimize the information freshness-aware task offloading in an air-ground integrated MEC system.
We formulate the interactions among the non-cooperative MUs across the infinite time-horizon as a stochastic game.
To approach the NE, each MU forms conjectures of the system states with the local observations of payment and computation service rate, which enables the transformation of the stochastic game into a single-agent MDP.
We then derive an online deep RL scheme that maintains two separate DQNs for each MU to approximate the Q-factor and the post-decision Q-factor.
Implementing the proposed deep RL scheme, each MU makes the decisions of channel auction, computation task offloading and input data packet scheduling only using the local information.
Numerical experiments confirm that compared with the four baselines, our scheme achieves a better tradeoff between the AoI and the energy consumption for all MUs in the system.



\begin{thebibliography}{29}

\bibitem{Mao17}
Y. Mao, C. You, J. Zhang, K. Huang and K. B. Letaief, ``A survey on mobile edge computing: The communication perspective,'' \emph{IEEE Commun. Surveys Tuts.}, vol. 19, no. 4, pp. 2322--2358, Q4 2017.

\bibitem{Wang20}
X. Wang, Y. Han, V. C. M. Leung, D. Niyato, X. Yan, and X. Chen, ``Convergence of edge computing and deep learning: A comprehensive survey,'' \emph{IEEE Commun. Surveys Tuts.}, vol. 22, no. 2, pp. 869--904, Q2 2020.

\bibitem{Wang18}
F. Wang, J. Xu, X. Wang, and S. Cui, ``Joint offloading and computing optimization in wireless powered mobile-edge computing systems,'' \emph{IEEE Trans. Wireless Commun.}, vol. 17, no. 3, pp. 1784--1797, Mar. 2018.

\bibitem{Liu17}
C.-F. Liu, M. Bennis, and H. V. Poor, ``Latency and reliability-aware task offloading and resource allocation for mobile edge computing,'' in \emph{Proc. IEEE GLOBECOM WKSHP}, Singapore, Dec. 2017.

\bibitem{Apos20}
P. A. Apostolopoulos, E. E. Tsiropoulou, and S. Papavassiliou, ``Risk-aware data offloading in multi-server multi-access edge computing environment,'' \emph{IEEE/ACM Trans. Netw.}, Early Access Article, 2020.

\bibitem{Chen19J}
X. Chen, H. Zhang, C. Wu, S. Mao, Y. Ji, and M. Bennis, ``Optimized computation offloading performance in virtual edge computing systems via deep reinforcement learning,'' \emph{ IEEE Internet Things J.}, vol. 6, no. 3, pp. 4005--4018, Jun. 2019.

\bibitem{He19}
X. He, R. Jin, and H. Dai, ``Deep PDS-Learning for Privacy-Aware Offloading in MEC-Enabled IoT,'' \emph{IEEE Internet Things J.}, vol. 6, no. 3, pp. 4547--4555, Jun. 2019.

\bibitem{Sun19}
Y. Sun, M. Peng, Y. Zhou, Y. Huang, and S. Mao, ``Application of machine learning in wireless networks: Key techniques and open issues,'' \emph{IEEE Commun. Surveys Tuts.}, vol. 21, no. 4, pp. 3072--3108, Q4 2019.

\bibitem{Chen19}
X. Chen, Z. Zhao, C. Wu, M. Bennis, H. Liu, Y. Ji, and H. Zhang, ``Multi-tenant cross-slice resource orchestration: A deep reinforcement learning approach,'' \emph{IEEE J. Sel. Areas Commun.}, vol. 37, no. 10, pp. 2377--2392, Oct. 2019.

\bibitem{Abd19}
M. A. Abd-Elmagid, A. Ferdowsi, H. S. Dhillon, and W. Saad, ``Deep reinforcement learning for minimizing age-of-information in UAV-assisted networks,'' in \emph{Proc. IEEE GLOBECOM}, Waikoloa, HI, Dec. 2019.

\bibitem{Zeng16M}
Y. Zeng, R. Zhang, and T. J. Lim, ``Wireless communications with unmanned aerial vehicles: Opportunities and challenges,'' \emph{IEEE Commun. Mag.}, vol. 54, no. 5, pp. 36--42, May 2016.

\bibitem{Moza19}
M. Mozaffari, W. Saad, M. Bennis, Y.-H. Nam, and M. Debbah, ``A tutorial on UAVs for wireless networks: Applications, challenges, and open problems,'' \emph{IEEE Commun. Surveys Tuts.}, vol. 21, no. 3, Q3 2019.

\bibitem{Amor20}
R. M. de Amorim, J. Wigard, I. Z. Kovacs, T. B. Sorensen, and P. E. Mogensen, ``Enabling cellular communication for aerial vehicles: Providing reliability for future applications,'' \emph{IEEE Veh. Technol. Mag.}, vol. 15, no. 2, pp. 129--135, Jun. 2020

\bibitem{Hu19}
X. Hu, K.-K. Wong, K. Yang, and Z. Zheng, ``UAV-assisted relaying and edge computing: Scheduling and trajectory optimization,'' \emph{IEEE Trans. Wireless Commun.}, vol. 18, no. 10, pp. 4738--4752, Oct. 2019.

\bibitem{Shan20}
B. Shang and L. Liu, ``Mobile edge computing in the sky: Energy optimization for air-ground integrated networks,'' \emph{IEEE Internet Things J.}, Early Access Article, 2020.

\bibitem{Ashe19}
A. Asheralieva and D. Niyato, ``Hierarchical game-theoretic and reinforcement learning framework for computational offloading in UAV-enabled mobile edge computing networks with multiple service providers,'' \emph{IEEE Internet Things J.}, vol. 6, no. 5, pp. 8753--8769, Oct. 2019.

\bibitem{ETSI18}
``Multi-access edge computing (MEC); Phase 2: Use cases and requirements,'' Oct. 2018, ETSI GS MEC 002 V2.1.1. [Online]. Available: \url{https://www.etsi.org/deliver/etsi_gs/MEC/001_099/002/02.01.01_60/gs_MEC002v020101p.pdf} [Accessed: 5 Jun. 2020].

\bibitem{Lian19}
Z. Liang, Y. Liu, T.-M. Lok, and K. Huang, ``Multiuser computation offloading and downloading for edge computing with virtualization,'' \emph{IEEE Trans. Wireless Commun.}, vol. 18, no. 9, pp. 4298--4311, Sep. 2019.

\bibitem{Wang20J}
X. Wang and L. Duan, ``Economic analysis of unmanned aerial vehicle (UAV) provided mobile services,'' \emph{IEEE Trans. Mobile Comput.}, Early Access Article, 2020.

\bibitem{Yate19}
R. D. Yates and S. K. Kaul, ``The age of information: Real-time status updating by multiple sources,'' \emph{IEEE Trans. Inf. Theory}, vol. 65, no. 3, pp. 1807--1827, Mar. 2019.

\bibitem{Yate20}
R. D. Yates, ``The age of information in networks: Moments, distributions, and sampling,'' \emph{IEEE Trans. Inf. Theory}, Early Access Article, 2020.

\bibitem{Kaul11}
S. Kaul, M. Gruteser, V. Rai, and J. Kenney, ``Minimizing age of information in vehicular networks,'' in \emph{Proc. IEEE SECON}, Salt Lake City, UT, Jun. 2011.

\bibitem{Chen19A}
X. Chen, C. Wu, T. Chen, H. Zhang, Z. Liu, Y. Zhang, and M. Bennis, ``Age of information-aware radio resource management in vehicular networks: A proactive deep reinforcement learning perspective,'' \emph{IEEE Trans. Wireless Commun.}, vol. 19, no. 4, pp. 2268--2281, Apr. 2020.

\bibitem{Abde18}
M. K. Abdel-Aziz, C. Liu, S. Samarakoon, M. Bennis, and W. Saad,``Ultra-reliable low-latency vehicular networks: Taming the age of information tail,''in \emph{Proc. IEEE GLOBECOM}, Abu Dhabi, UAE, Dec. 2018.

\bibitem{Zhon19}
J. Zhong, W. Zhang, R. D. Yates, A. Garnaev, and Y. Zhang, ``Age-aware scheduling for asynchronous arriving jobs in edge applications,'' in \emph{Proc. IEEE INFOCOM WKSHP}, Paris, France, Apr.--May 2019.

\bibitem{Xu20}
C. Xu, H. H. Yang, X. Wang, and T. Q. S. Quek, ``Optimizing information freshness in computing-enabled IoT networks,'' \emph{IEEE Internet Things J.}, vol. 7, no. 2, pp. 971--985, Feb. 2020.

\bibitem{Kuan20}
Q. Kuang, J. Gong, X. Chen, and X. Ma, ``Analysis on computation-intensive status update in mobile edge computing,'' \emph{IEEE Trans. Veh. Technol.}, vol. 69, no. 4, pp. 4353--4366, Apr. 2020.

\bibitem{Ji07}
Z. Ji and K. J. R. Liu, ``Dynamic spectrum sharing: A game theoretical overview,'' \emph{IEEE Commun. Mag.}, vol. 45, no. 5, pp. 88--94, May 2007.

\bibitem{Edel07}
B. Edelman, M. Ostrovsky, and M. Schwarz, ``Internet advertising and the generalized second-price auction: Selling billions of dollars worth of keywords,'' \emph{Am. Econ. Rev.}, vol. 97, no. 1, pp. 242--259, Mar. 2007.

\bibitem{Hass16}
H. van Hasselt, A. Guez, and D. Silver, ``Deep reinforcement learning with double Q-learning,'' in \emph{Proc. AAAI}, Phoenix, AZ, Feb. 2016.

\bibitem{Mnih16}
V. Mnih, A. P. Badia, M. Mirza, A. Graves, T. Harley, T. P. Lillicrap, D. Silver, and K. Kavukcuoglu, ``Asynchronous methods for deep reinforcement learning,'' in \emph{Proc. ICML}, New York City, NY, Jun. 2016.

\bibitem{Liu19}
X. Liu, Y. Liu, Y. Chen, and L. Hanzo, ``Trajectory design and power control for multi-UAV assisted wireless networks: A machine learning approach,'' \emph{IEEE Trans. Veh. Technol.}, vol. 68, no. 8, pp. 7957--7969, Aug. 2019.

\bibitem{Rich98}
R. S. Sutton and A. G. Barto, \emph{Reinforcement Learning: An Introduction}. Cambridge, MA: MIT Press, 1998.

\bibitem{Wan13}
Y. Wan, K. Namuduri, Y. Zhou, and S. Fu, ``A smooth-turn mobility model for airborne networks,'' \emph{IEEE Trans. Veh. Technol.}, vol. 62, no. 7, pp. 3359--3370, Sep. 2013.

\bibitem{Xi19}
X. Xi, X. Cao, P. Yang, Z. Xiao, and D. Wu, ``Efficient and fair network selection for integrated cellular and drone-cell networks,'' \emph{IEEE Trans. Veh. Technol.}, vol. 68, no. 1, pp. 923--937, Jan. 2019.

\bibitem{Amer20}
R. Amer, W. Saad, and N. Marchetti, ``Mobility in the sky: Performance and mobility analysis for cellular-connected UAVs,'' IEEE Trans. Commun., vol. 68, no. 5, pp. 3229--3246, May 2020.

\bibitem{Burd96}
T. D. Burd and R. W. Brodersen, ``Processor design for portable systems,'' \emph{J. VLSI Signal Process. Syst.}, vol. 13, no. 2–3, pp. 203--221, Aug. 1996.

\bibitem{Chen16}
X. Chen, L. Jiao, W. Li, and X. Fu, ``Efficient multi-user computation offloading for mobile-edge cloud computing,'' \emph{IEEE/ACM Trans. Netw.}, vol. 24, no. 5, pp. 2795--2808, Oct. 2016.

\bibitem{Fied10}
M. Fiedler, T. Hossfeld, and P. Tran-Gia, ``A generic quantitative relationship between quality of experience and quality of service,'' \emph{IEEE Netw.}, vol. 24, no. 2, pp. 36--41, Mar./Apr. 2010.

\bibitem{Adel08}
D. Adelman and A. J. Mersereau, ``Relaxations of weakly coupled stochastic dynamic programs,'' \emph{Oper. Res.}, vol. 56, no. 3, pp. 712--727, Jan. 2008.

\bibitem{Fink64}
A. M. Fink, ``Equilibrium in a stochastic $n$-person game,'' \emph{J. Sci. Hiroshima Univ. Ser. A-I}, vol. 28, pp. 89--93, 1964.

\bibitem{Chen1801}
X. Chen, Z. Han, H. Zhang, G. Xue, Y. Xiao, and M. Bennis, ``Wireless resource scheduling in virtualized radio access networks using stochastic learning,'' \emph{IEEE Trans. Mobile Comput.}, vol. 17, no. 4, pp. 961--974, Apr. 2018.

\bibitem{Kroe16}
C. Kroer and T. Sandholm, ``Imperfect-recall abstractions with bounds in games,'' in \emph{Proc. ACM EC}, Maastricht, the Netherlands, Jul. 2016.

\bibitem{Salo08}
N. Salodkar, A. Bhorkar, A. Karandikar, and V. S. Borkar, ``An on-line learning algorithm for energy efficient delay constrained scheduling over a fading channel,'' \emph{IEEE J. Sel. Areas Commun.}, vol. 26, no. 4, pp. 732--742, May 2008.

\bibitem{Chen18S}
X. Chen, C. Wu, M. Bennis, Z. Zhao, and Z. Han, ``Learning to entangle radio resources in vehicular communications: An oblivious game-theoretic perspective,'' \emph{IEEE Trans. Veh. Technol.}, vol. 68, no. 5, pp. 4262--4274, May 2019.


\bibitem{Mast13}
N. Mastronarde and M. van der Schaar, ``Joint physical-layer and system-level power management for delay-sensitive wireless communications,'' \emph{IEEE Trans. Mobile Comput.}, vol. 12, no. 4, pp. 694--709, Apr. 2013.

\bibitem{Appl17}
Apple, ``The future is here: iPhone X'', 2017. [Online]. Available: \url{https://www.apple.com/newsroom/2017/09/the-future-is-here-iphone-x/} [Accessed: 16 Feb. 2020].

\bibitem{Mnih15}
V. Mnih, K. Kavukcuoglu, D. Silver, A. A. Rusu, J. Veness, M. G. Bellemare, A. Graves, M. Riedmiller, A. K. Fidjeland, G. Ostrovski, S. Petersen, C. Beattie, A. Sadik, I. Antonoglou, H. King, D. Kumaran, D. Wierstra, S. Legg, and D. Hassabis, ``Human-level control through deep reinforcement learning,'' \emph{Nature}, vol. 518, no. 7540, pp. 529--533, Feb. 2015.

\bibitem{Ram10}
J. A. Ram\'{\i}rez-Hern\'{a}ndez and E. Fernandez, ``Optimization of preventive maintenance scheduling in semiconductor manufacturing models using a simulation-based approximate dynamic programming approach,'' in \emph{IEEE CDC}, Atlanta, GA, Dec. 2010.

\bibitem{Wang19}
Y. Wang and D. R. Jiang, ``Structured actor-critic for managing public health points-of-dispensing,'' 2019.  [Online]. Available: \url{https://arxiv.org/pdf/1806.02490.pdf} [Accessed: 18 Feb. 2020].

\bibitem{Lin92}
L.-J. Lin, ``Reinforcement learning for robots using neural networks,'' Carnegie Mellon University, 1992.

\bibitem{Abad16}
M. Abadi, P. Barham, J. Chen, Z. Chen, A. Davis, J. Dean, M. Devin, S. Ghemawat, G. Irving, M. Isard, M. Kudlur, J. Levenberg, R. Monga, S. Moore, D. G. Murray, B. Steiner, P. Tucker, V. Vasudevan, P. Warden, M. Wicke, Y. Yu, and X. Zheng, ``Tensorflow: A system for large-scale machine learning,'' in \emph{Proc. OSDI}, Savannah, GA, Nov. 2016.

\bibitem{Zeng16}
Y. Zeng, R. Zhang, and T. J. Lim, ``Throughput maximization for UAV-enabled mobile relaying systems,'' \emph{IEEE Trans. Commun.}, vol. 64, no. 12, pp. 4983--4996, Dec. 2016.

\bibitem{Nair10}
V. Nair and G. E. Hinton, ``Rectified linear units improve restricted boltzmann machines,'' in \emph{Proc. ICML}, Haifa, Israel, Jun. 2010.

\bibitem{King15}
D. P. Kingma and J. Ba, ``Adam: A Method for Stochastic Optimization,'' in \emph{Proc. ICLR}, San Diego, CA, May 2015.

\end{thebibliography}
\end{document}